\documentclass[conf]{new-aiaa} 
\usepackage[utf8]{inputenc}
\usepackage{textcomp}

\usepackage{graphicx}
\usepackage{color}
\usepackage{amsmath, amsfonts}
\usepackage{amssymb}
\usepackage[version=4]{mhchem}
\usepackage{siunitx}
\usepackage{longtable,tabularx}

\usepackage{bm}
\usepackage{subfigure}
\usepackage{wasysym}
\usepackage{algorithm}
\usepackage{multirow}
 
\usepackage{url}
\usepackage{braket}
\usepackage{fancyhdr}
\newif\ifmajorrev
\majorrevfalse 
\newcommand{\majorblue}[1]{\ifmajorrev\textcolor{blue}{#1}\else{#1}\fi}
\newcommand{\majorred}[1]{\ifmajorrev\textcolor{red}{#1}\else{#1}\fi}

\newif\ifminorrev
\minorrevfalse 
\newcommand{\minorblue}[1]{\ifminorrev\textcolor{blue}{#1}\else{#1}\fi}

\newcommand{\destiny}{DESTINY$^+$}

\title{Asteroid Flyby Cycler Trajectory Design Using Deep Neural Networks}

\author{Naoya Ozaki\footnote{Ph.D., Assistant Professor, Department of Spacecraft Engineering, Japan Aerospace Exploration Agency, Kanagawa 252-5210, Japan; ozaki.naoya@jaxa.jp, and Member AIAA.}}
\affil{Japan Aerospace Exploration Agency, Sagamihara, Kanagawa, 252-5210, Japan}

\author{Kanta Yanagida\footnote{Ph.D. Student, Department of Aeronautics and Astronautics, the University of Tokyo, 7-3-1, Hongo, Bunkyo-ku, Tokyo, 113-8656, Japan.},
Takuya Chikazawa\footnote{Ph.D. Student, Department of Advanced Energy, the University of Tokyo, 7-3-1, Hongo, Bunkyo-ku, Tokyo, 113-8654, Japan.}}
\affil{The University of Tokyo, Tokyo 113-8654, Japan}

\author{Nishanth Pushparaj\footnote{Ph.D. Student, Department of Space and Astronautical Science, The Graduate University for Advanced Studies, SOKENDAI, Sagamihara, Kanagawa, 252-5210, Japan.}}
\affil{SOKENDAI, Sagamihara, Kanagawa, 252-5210, Japan}

\author{Naoya Takeishi\footnote{Ph.D., Postdoctoral Researcher, Geneva School of Business Administration, University of Applied Sciences and Arts Western Switzerland, 1227 Carouge, Switzerland; naoya.takeishi@hesge.ch.}}
\affil{University of Applied Sciences and Arts Western Switzerland, 1227 Carouge, Switzerland}

\author{Ryuki Hyodo\footnote{Ph.D., International Top Young Research Fellow, Department of Solar System Sciences, Japan Aerospace Exploration Agency, Kanagawa 252-5210, Japan.}}
\affil{Japan Aerospace Exploration Agency, Sagamihara, Kanagawa, 252-5210, Japan}

\begin{document}

\maketitle

\begin{abstract}
Asteroid exploration has been attracting more attention in recent years. Nevertheless, we have just visited tens of asteroids while we have discovered more than one million bodies. As our current observation and knowledge should be biased, it is essential to explore multiple asteroids directly to better understand the remains of planetary building materials. One of the mission design solutions is utilizing asteroid flyby cycler trajectories with multiple Earth gravity assists. 
An asteroid flyby cycler trajectory design problem is a subclass of global trajectory optimization problems with multiple flybys, involving a trajectory optimization problem for a given flyby sequence and a combinatorial optimization problem to decide the sequence of the flybys. As the number of flyby bodies grows, the computation time of this optimization problem expands maliciously. 
This paper presents a new method to design asteroid flyby cycler trajectories utilizing a surrogate model constructed by deep neural networks approximating trajectory optimization results. Since one of the bottlenecks of machine learning approaches is \majorblue{the \minorblue{heavy} computation time} to generate massive trajectory databases, we propose an efficient database generation strategy by introducing pseudo-asteroids satisfying the Karush-Kuhn-Tucker conditions. 
The numerical result applied to JAXA's \destiny mission shows that the proposed method \majorblue{is practically applicable to space mission design and} can significantly reduce the computational time for searching asteroid flyby sequences.
\end{abstract}

\section*{Nomenclature}

{\renewcommand\arraystretch{1.0}
\noindent\begin{longtable*}{@{}l @{\quad=\quad} l@{}}
$\textbf{\oe}$  & Keplerian orbital elements\\
$a$ & semi-major axis\majorblue{, km}\\ 
$e$ & eccentricity\\
$i$ & inclination\majorblue{, rad}\\ 
$\Omega$ & longitude of the ascending node\majorblue{, rad}\\ 
$\Delta \Omega$ & difference of the asteroid $\Omega$ from the Earth departure longitude\majorblue{, rad}\\ 
$\omega$ & argument of perihelion\majorblue{, rad}\\ 
$M_{\textrm{epoch}}$ & mean anomaly at the epoch\majorblue{, rad}\\ 
$\lambda$ & longitude\majorblue{, rad}\\ 
$\bm{r}$ & position \majorblue{vector, km}\\ 
$\bm{v}$ & velocity \majorblue{vector, km/s}\\ 
$\bm{v}_{\infty}$ & hyperbolic excess velocity \majorblue{vector} with respect to the Earth\majorblue{, km/s}\\ 
$m$ & integer number of Earth full revolution; $m\in\mathbb{Z}_{\geq 0}$\\
$n$ & integer number of spacecraft full revolution; $n\in\mathbb{Z}_{\geq 0}$\\
$\alpha$ & pump angle\majorblue{, rad}; $\alpha\in[0,\pi]$\\ 
$\kappa$ & crank angle\majorblue{, rad}; $\kappa\in[0,2\pi)$\\ 
\majorblue{$T$} & time of flight\majorblue{, sec}\\ 
$\Delta v$ & magnitude of impulsive velocity change\majorblue{, km/s}\\ 
\majorblue{$\mu$} & gravitational parameter\majorblue{, km$^3$/s$^2$}\\ 
\multicolumn{2}{@{}l}{Subscripts}\\
UB & upper bound\\
LB & lower bound\\
in & incoming\\
out & outgoing\\
sc & spacecraft\\
$\odot$ & sun\\
$\oplus$ & Earth\\
$\star$ & asteroid
\end{longtable*}}

%
%
%
%
\section{Introduction}

%
%
%

Multiple asteroid flyby missions place new scientific constraints on, e.g., chemical, spectral, and geomorphological diversities among asteroids by in-situ surface observations. One of the mission design solutions to visit many asteroids is using asteroid flyby cycler trajectories\cite{Englander2019}, a special case of free-return cyclers. Free-return cyclers are periodic trajectories that shuttle a spacecraft between two or more celestial bodies and are typically used in the Sun-Earth-Mars system\cite{Byrnes1993, Russell2005a} and planetary moon systems\cite{Russell2009, Campagnola2019}. Unlike typical free-return cyclers, asteroid flyby cyclers do not require the periodicity of all associated bodies' relative geometry because the spacecraft targets different bodies after each Earth gravity assist. An asteroid flyby cycler trajectory design problem \majorblue{addressed in this paper} is a subclass of global trajectory optimization problems with multiple flybys \majorblue{and gravity assists}, which has attracted many mission designers as one of the most challenging problems. 

%
%
%

Global trajectory optimization problems with multiple flybys essentially involve two optimization problems: a nonlinear optimal control problem that optimizes the trajectory for a given flyby sequence and a combinatorial optimization problem to choose the sequence of the flybys. Recent studies have proposed global trajectory optimization methods utilizing population-based metaheuristics\cite{Englander2012, Englander2017, Petropoulos2018}\majorblue{, a binary tree search\cite{Carlo2018}, and indirect methods\cite{Olympio2011, Grigoriev2013}.} Englander et al. have applied their evolutionary algorithm-based method to NASA's Lucy mission, which visits multiple Trojan asteroids\cite{Englander2019}. \majorblue{Sánchez Cuartielles et al. and Bowles et al. have studied multiple-flyby trajectory design methods via genetic algorithm for the CASTAway mission\cite{Cuartielles2016, Bowles2018}.} Although metaheuristic approaches are successfully implemented for practical missions, the approaches face computational difficulties when the number of target bodies grows. We need further studies to search for target asteroids among the million of them comprehensively. \majorblue{As a post-analysis discussion,} Englander et al. have also remarked a trend among orbital elements $\omega$ and $\Omega$ of accessible asteroids\cite{Englander2019}. This trend indicates that \majorblue{embedding a well-designed} surrogate model \majorblue{into a global trajectory optimization process enables us to} search for accessible asteroids efficiently.

In machine learning and optimization theory, researchers have studied efficient global optimization algorithms utilizing surrogate models\cite{Haftka2016}. \majorred{A surrogate model is a black-box model that approximates the relationship between inputs and outputs rather than calculates it directly. We can rapidly obtain the global optimal solution by replacing time-consuming trajectory optimization with a less time-consuming surrogate model constructed by a neural network.} One of the most successful examples is Bayesian optimization\cite{Brochu2010}, which commonly uses a Gaussian process as a surrogate model. This method is widely used to tune hyperparameters of machine learning\cite{Snoek2012}. Some of the other methods implement surrogate models in an evolutionary algorithm\cite{Jin2011} and particle swarm optimization\cite{Parno2012}. In recent years, astrodynamics researchers have started applying machine learning techniques to space mission design\majorblue{\cite{Yang2021, Yang_inpress, Izzo2018}}, and some of them have presented surrogate-assisted global trajectory optimization methods\majorblue{\cite{Hennes2017, Mereta, Shang2017, Shang2018, Viavattene_inpress, Li2019, Zhu2019}}. Their surrogate models, made by classical regression method\cite{Hennes2017, Mereta}, Gaussian processes\majorblue{\cite{Shang2017,Shang2018}}, Deep Neural Networks (DNNs)\cite{Viavattene_inpress, Li2019, Zhu2019, Izzo2021}, approximate the cost function of the nonlinear trajectory optimization problem and let us evaluate the cost quickly without solving the trajectory optimization problems. Although the DNN-based method accurately approximates the actual cost function\majorblue{, for example, multiple rendezvous missions with near-Earth asteroids (NEAs)\cite{Viavattene_inpress},} its training requires a computationally expensive massive database (e.g., hundreds of thousands of optimal trajectories). Izzo and \"{O}zt\"{u}rk\cite{Izzo2021} have studied an innovative approach that generates a massive database efficiently without solving trajectory optimization problems directly\majorred{, yet this prior study is only applicable to low-thrust controllers for two-point boundary value problems such as the Earth-Venus transfer.}

%
%
%

This paper presents a novel method to design asteroid flyby cycler trajectories utilizing a surrogate model constructed by DNNs. \majorblue{Our approach} builds the surrogate model of nonlinear trajectory optimization and \majorblue{then} efficiently searches for flyby sequences using the tree search method with the surrogate model. \majorred{Our first contribution is to bring the surrogate-based approach to a practical flyby trajectory design by improving the prediction accuracy of our surrogate model by utilizing astrodynamics knowledge, such as free-return trajectories\cite{Russell2005, Russell2009, Prussing2000, Shen2004} and Lambert's problem\cite{Izzo2015}. The second contribution is to allow the resulting surrogate model to be reusable for different mission scenarios (different Earth departure epoch, hyperbolic excess velocity, and target asteroids) without re-training DNNs by normalizing the dataset with the longitude at the Earth departure epoch. The third contribution is to establish an efficient database generation strategy that can produce multiple datasets from a single trajectory optimization result by introducing pseudo-asteroids while maintaining the optimization condition, the Karush–Kuhn–Tucker (KKT) condition. This strategy can amplify the size of the trajectory optimization database by one order of magnitude by solving a simple algebraic equation. Finally, we search for optimal flyby sequence by a beam search method\cite{Izzo2016} where the heuristic cost is quickly estimated by utilizing the surrogate model. Our proposed approach is tested in the numerical application to JAXA's \destiny mission\cite{Ozaki2022} performing multiple asteroid flybys, including (3200) Phaethon.}




%
%
\section{Background}\label{sec:previous_works}

\begin{figure}[htb]
\begin{center}
\includegraphics[width=0.45\hsize]{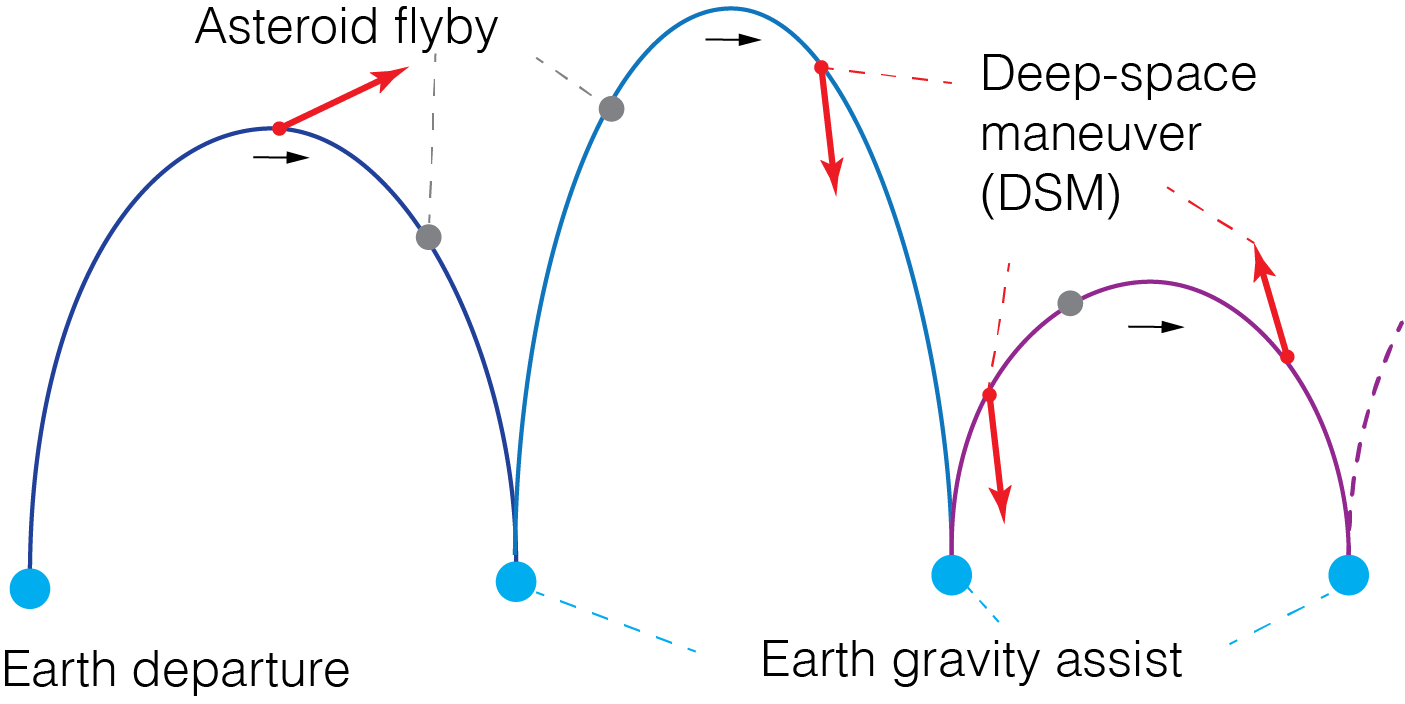}
\caption{\label{f:multiple_asteroid_flyby_trajectory}Asteroid flyby cycler.}
\end{center}
\end{figure}

This section provides the necessary definition of the dynamics, free-return trajectories, and the trajectory optimization problem, including the assumptions and the notation. Our goal is to fly by as many scientifically interesting asteroids as possible within the limited spacecraft capability (propellant and lifetime). In particular, we assume that the fuel constraints are severe, while the requirements on lifetime are relatively loose. To this end, we utilize Earth gravity-assist maneuvers repeatedly, achieved by the Earth free-return trajectories, as shown in Fig.\ref{f:multiple_asteroid_flyby_trajectory}. This approach reduces the total fuel consumption significantly with a relatively long time of flight.


%
%
\subsection{Dynamics and Models}

We consider that the dynamics of spacecraft, asteroids, and the Earth are governed only by the sun's gravity, and the Earth is moving in a circular orbit on the ecliptic plane\majorblue{, known as the zero-radius sphere-of-influence patched-conics model\cite{Vasile2006, Ozaki2022a}}. The Earth gravity assist is modeled as an instantaneous velocity change at the intersection of the spacecraft and the Earth's orbits. We describe the spacecraft heliocentric state via Cartesian elements \majorblue{$\bm{x}=\left[\bm{r}^{\top}, \bm{v}^{\top}\right]^{\top}$}. The equation of motion of the spacecraft is described as follows:
\begin{equation}
    \frac{d}{dt}\begin{bmatrix}
    \bm{r}\\
    \bm{v}
    \end{bmatrix}= \begin{bmatrix}
    \bm{v}\\
    -\frac{\majorblue{\mu}_{\odot}}{\|\bm{r}\|^3}\bm{r}
    \end{bmatrix}
\end{equation}

We formulate the trajectory optimization problem via the multiple gravity-assists with one deep-space maneuver (MGA-1DSM) model\cite{Vinko2008, Vasile2006, Englander2012}, where a single deep-space maneuver can be performed at any point along the trajectory between each consecutive flybys. Although we demonstrate our algorithm using the MGA-1DSM model, this algorithm can be potentially extended to low-thrust trajectory optimization if the computational time allows.


%
%
\subsection{Free-return Trajectories}

\begin{figure}[htb]
\begin{center}
\includegraphics[width=0.8\hsize]{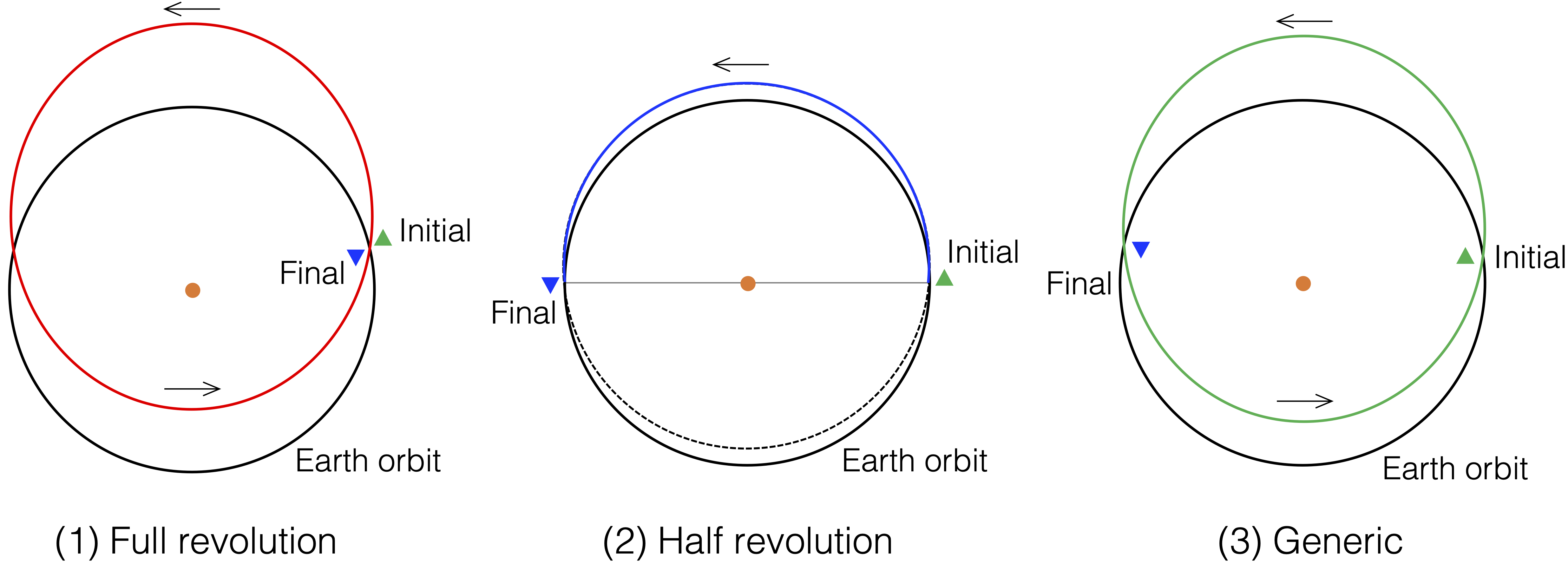}
\caption{\label{f:type_freereturn}Three category of free-return trajectories.}
\end{center}
\end{figure}

\begin{figure}[htb!]
\begin{center}
\includegraphics[width=0.5\hsize]{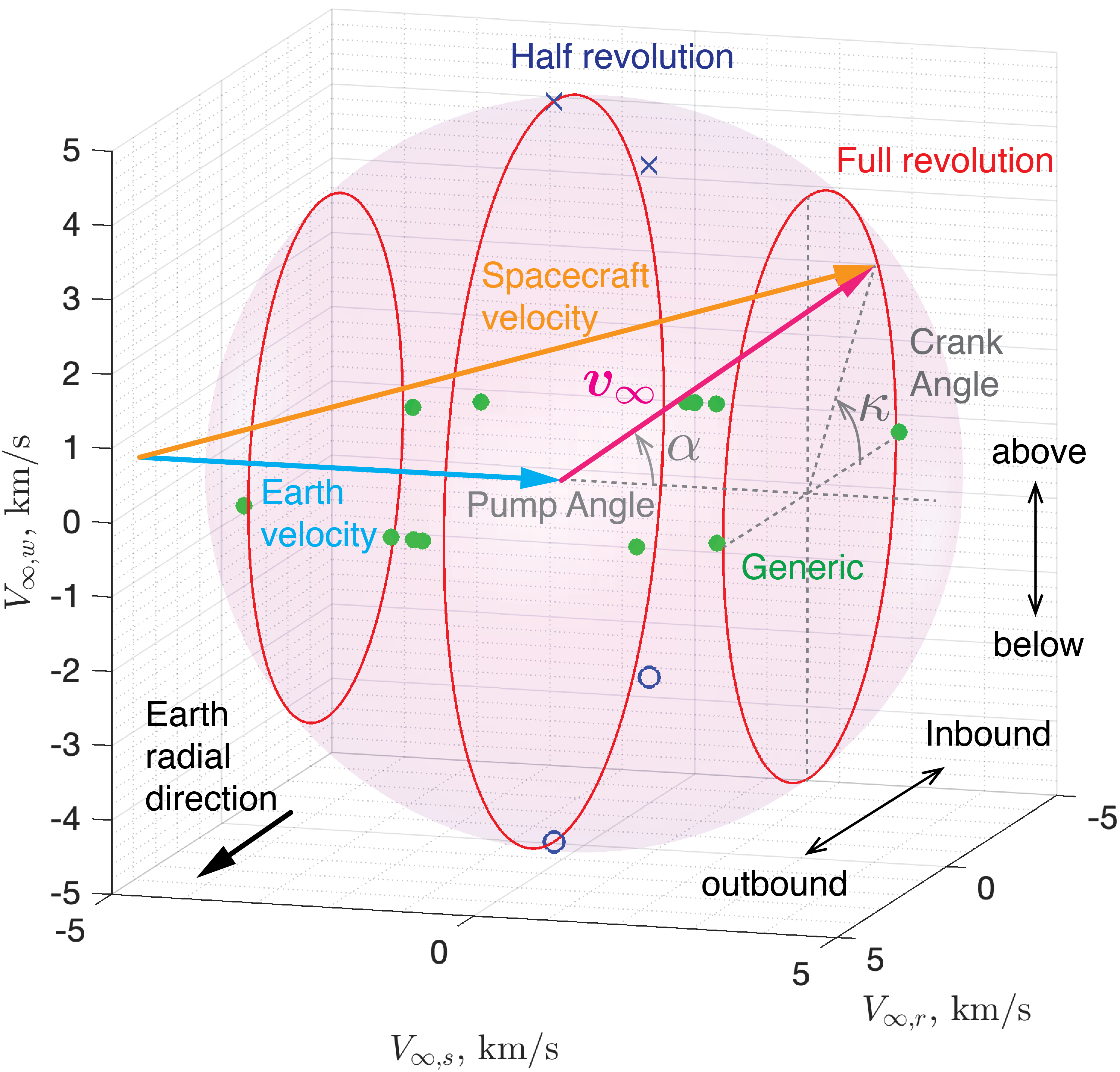}
\caption{\label{f:vinf_globe}The v-infinity globe of Earth free-return trajectories ($V_{\infty}=5$km/s and $m\leq 3$)\majorblue{, where} $s$ axis along the velocity vector of the flyby body, $w$ axis along the angular momentum vector of the flyby body, and $r$ axis completes the right-handed system.}
\end{center}
\end{figure}

Our studies investigate asteroid flyby cyclers using Earth free-return trajectories and patching them together with Earth gravity-assist maneuvers. Russell et al.\cite{Russell2005, Russell2009} have proposed a systematic method to identify all feasible free-return trajectories under the assumption that the flyby body is moving in a circular orbit. As per their approach, the free-return trajectories fall into three categories shown in Fig.\ref{f:type_freereturn}: 1) full-revolution transfers, 2) half-revolution transfers, and 3) generic transfers. The free-return trajectory can be uniquely determined when the parameters shown in Table \ref{tab:freereturn_parameters} are fixed.  Introducing the v-infinity globe\cite{Strange2008}, we can plot all free-return trajectories as points on the globe as shown in Fig.\ref{f:vinf_globe}. In Fig.\ref{f:vinf_globe}, the red large circles represent the v-infinity direction of the full revolution free-return trajectories; the blue markers indicate the half revolution trajectories; the green dots indicate the generic free-return trajectories. Details for calculating all free-return trajectories can be found in Ref.\cite{Russell2005, Russell2009, Prussing2000, Shen2004}.

\begin{table}[hbt!]
\caption{\label{tab:freereturn_parameters} Parameters determining free-return trajectories}
\centering
\begin{tabular}{lcc}
\hline
Type & Continuous & Discrete \\ \hline
full& $v_{\infty}\in\mathbb{R}_{\geq0}$, $\kappa\in[0, 2\pi)$ & $m, n\in\mathbb{Z}_{\geq 0}$\\
half& $v_{\infty}\in\mathbb{R}_{\geq0}$ & $m, n\in\mathbb{Z}_{\geq 0}$, \{inbound, outbound\}, \{above, below\} \\
generic& $v_{\infty}\in\mathbb{R}_{\geq0}$ & $m, n\in\mathbb{Z}_{\geq 0}$, \{inbound, outbound\}, \{direct, retrograde\} \\
\hline
\end{tabular}
\end{table}


%
%
\subsection{Asteroid Flyby Cycler Trajectory Design Problem}

\majorred{We formulate the asteroid flyby cycler trajectory design problem by dividing the whole problem into smaller subproblems\cite{Izzo2016}.} One of them is an Earth-asteroid-Earth trajectory optimization problem under the MGA-1DSM models to construct the Earth-asteroid-Earth block. The other subproblem is to search the sequence of target asteroids via tree searches. \majorred{Although this greedy formulation only yields a series of optimal trajectories that approximate the global optimal trajectory, this simplification allows us to demonstrate surrogate-based flyby trajectory design with practical resources.} We also insert Earth free-return transfers without visiting any asteroids if needed (mainly for changing Earth flyby longitude). Visiting more than one asteroid between each Earth-to-Earth leg is out of the main scope of this research. Once the sequence of asteroid flybys is determined, we optimize the patched trajectory and evaluate the total $\Delta V$ consumption and the total time of flight.

%
%
\subsubsection{Earth-Asteroid-Earth Block}

\begin{figure}[htb]
\begin{center}
\includegraphics[width=0.5\hsize]{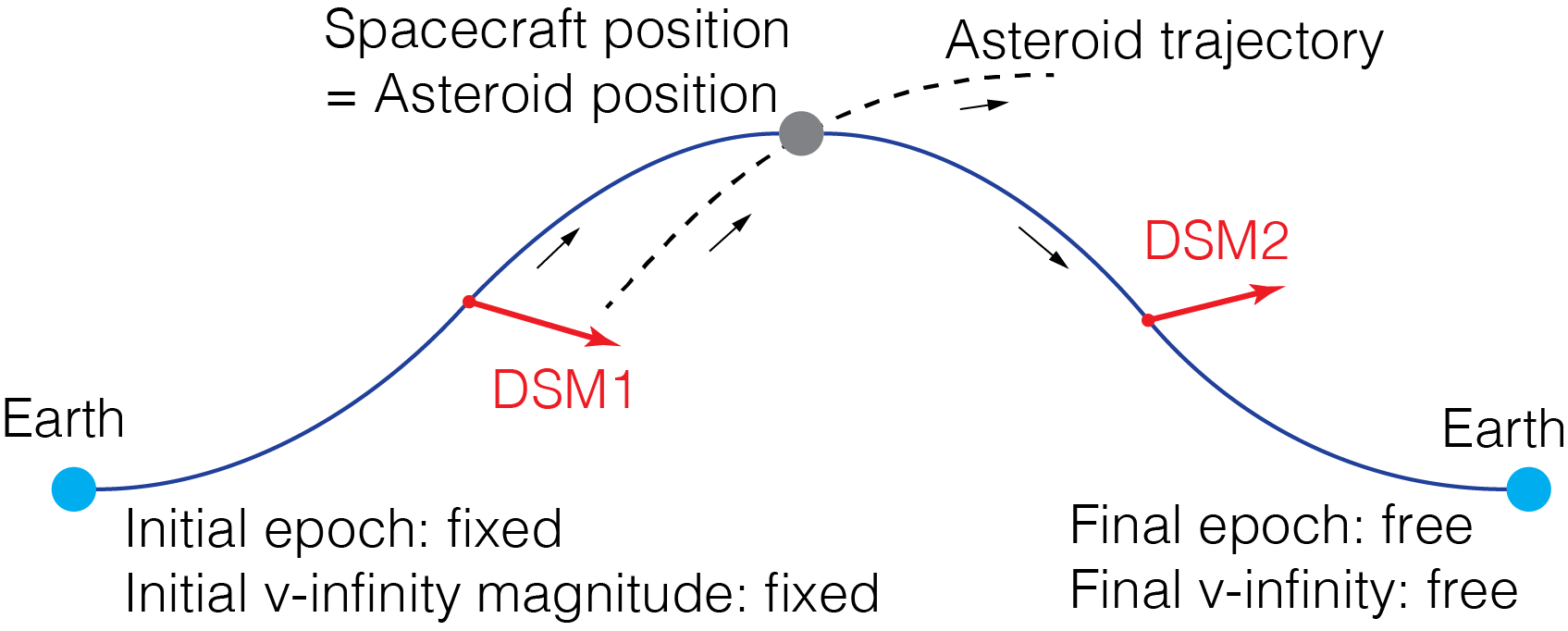}
\caption{\label{f:eae_block}Definition of Earth-asteroid-Earth trajectory optimization problem.}
\end{center}
\end{figure}

We formulate the Earth-asteroid-Earth transfer problems by two consecutive MGA-1DSM models, where the gravities of asteroids are ignored, as shown in Fig.\ref{f:eae_block}. The main task of this subproblem is to calculate the optimal Earth-asteroid-Earth transfers that minimize the total $\Delta V$ (the sum of DSM1 and DSM2). Initial epoch $t_{0}$ and initial v-infinity magnitude $v_{\infty, 0}$ are fixed; final epoch $t_f$ and final v-infinity $v_{\infty, f}$ are free. The spacecraft performs an asteroid flyby during the Earth-to-Earth transfer; that is, the positions of two bodies match at $t_{\star}\in(t_0, t_f)\subset\mathbb{R}$. 

\begin{figure}[htb]
\begin{center}
\includegraphics[width=0.4\hsize]{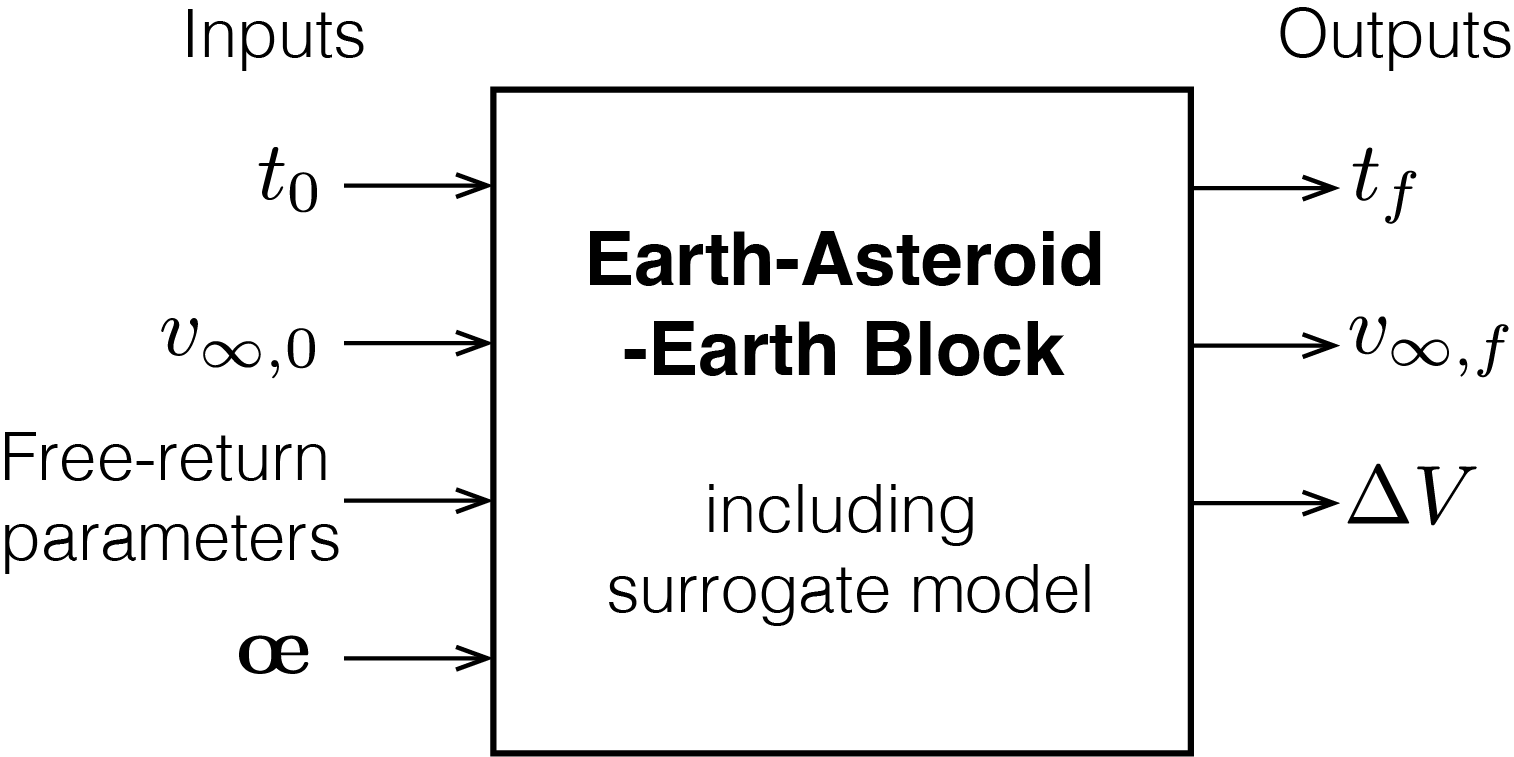}
\caption{\label{f:innerloop}Inputs and outputs of the Earth-asteroid-Earth block.}
\end{center}
\end{figure}

The inputs of this block are $t_0$, $v_{\infty, 0}$, the \majorblue{parameters} of free-return trajectory, and the orbital elements of the target asteroid $\textbf{\oe} = [a,e,i,\Omega,\omega,M_{t_\textrm{\oe}}]$. As the outputs of this problem, we obtain the total $\Delta V$ consumption, \majorblue{$t_f$}, and $v_{\infty, f}$. Figure \ref{f:innerloop} illustrates the relationship between the inputs and outputs of the Earth-asteroid-Earth transfer problem.


%
%
\subsubsection{Tree Searches of Flyby Sequence}

\begin{figure}[htb]
\begin{center}
\includegraphics[width=0.7\hsize]{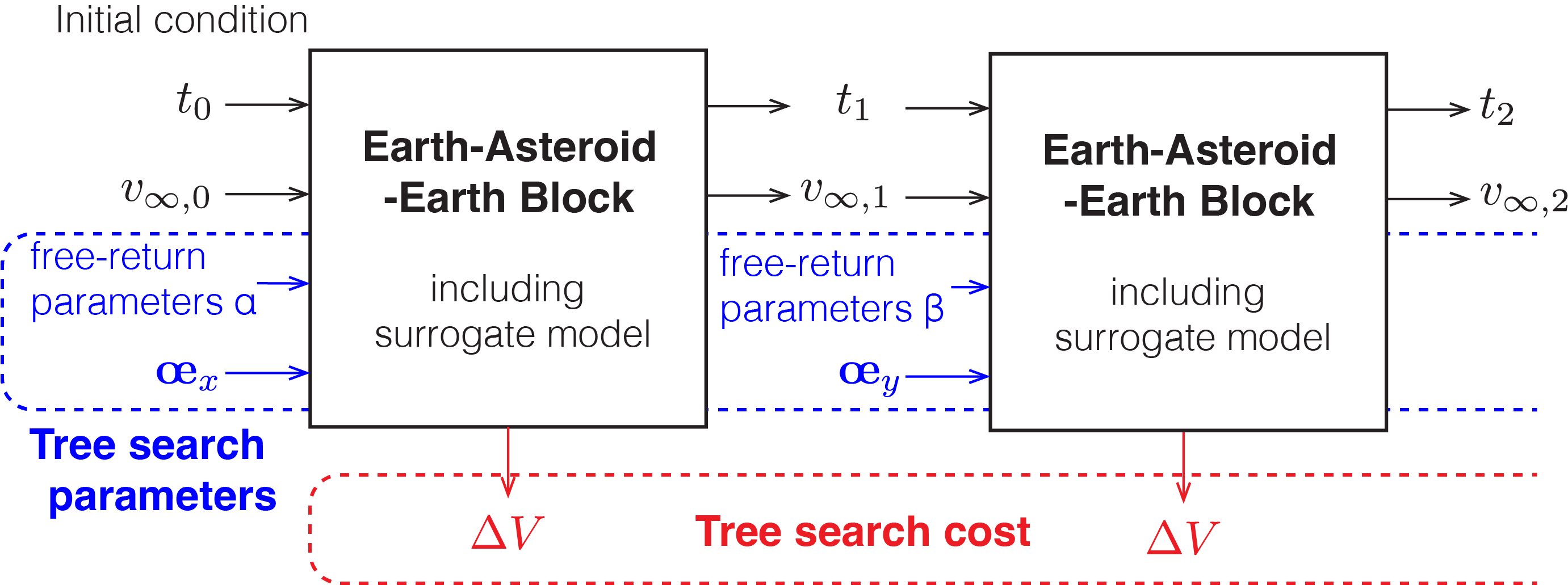}
\caption{\label{f:outerloop}Flyby sequence tree search using Earth-asteroid-Earth block.}
\end{center}
\end{figure}

Using the result of the Earth-asteroid-Earth block, we search for the good sequences of \majorblue{free-return parameters and} target asteroids via tree search methods\cite{Izzo2016}. \majorblue{The tree search parameters are the free-return parameters $(m, n, \textrm{type})$ and the orbital elements of the asteroids \textbf{\oe}. Figure \ref{f:outerloop} illustrates an example sequence where we first select $\alpha$ as the free-return parameter and \textbf{\oe}$_x$ as the orbital element and then select $\beta$ as the free-return parameter and \textbf{\oe}$_y$ as the orbital element.} The cost of the flyby sequences depends on the solution of the Earth-asteroid-Earth block, whereas the inputs of the Earth-asteroid-Earth block are given when the flyby sequence is determined. Hence, without the surrogate model, we need to solve trajectory optimization problems to evaluate the cost of the tree nodes.

%
%


\section{Asteroid Flyby Cycler Trajectory Design via Deep Neural Networks}\label{sec:extended}

\begin{figure}[htb]
\begin{center}
\includegraphics[width=\hsize]{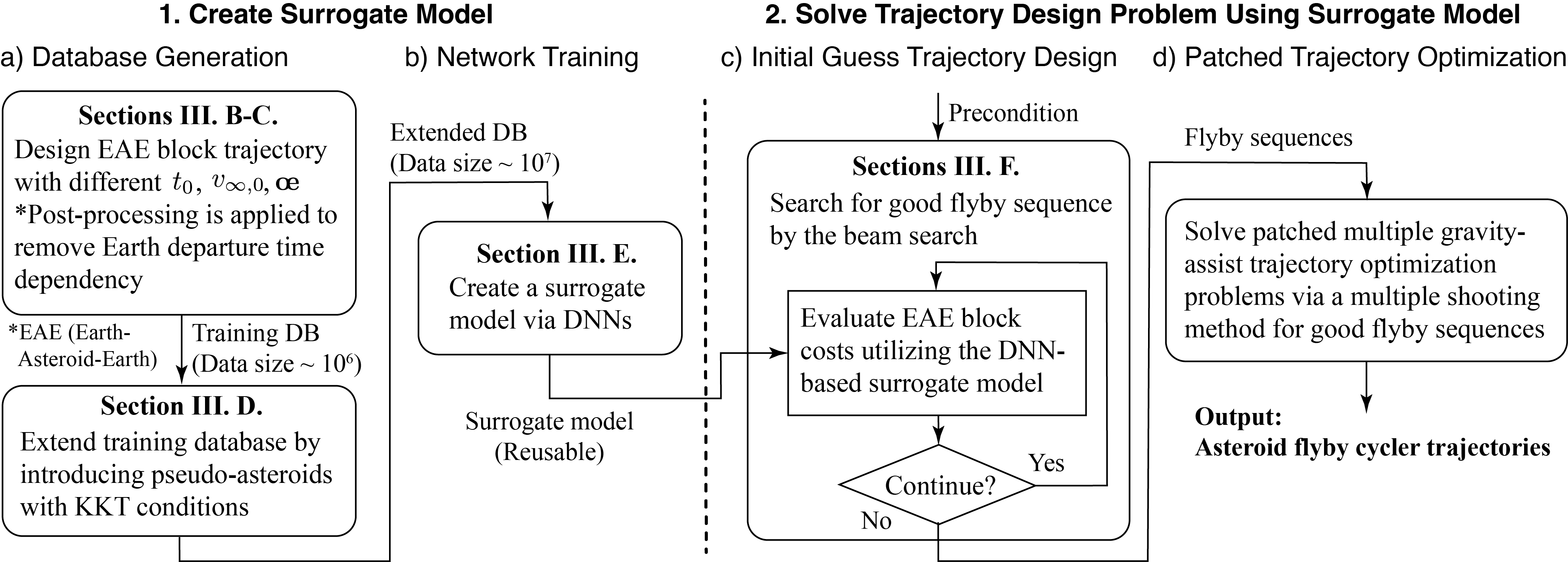}
\caption{\label{f:trajectory_design_step}\majorred{Surrogate-based asteroid flyby cycler trajectory design procedure.}}
\end{center}
\end{figure}

This section presents a novel asteroid flyby cycler design method using the surrogate model constructed via DNNs. \majorred{Figure \ref{f:trajectory_design_step} illustrates the proposed trajectory design procedure and indicates the corresponding sections. We create the surrogate model of the Earth-asteroid-Earth block that allows us to quickly estimate the cost of each transfer between asteroids without solving trajectory optimization problems during the tree search.} Since the DNN-based methods require gigantic databases, we propose an efficient database generation strategy by introducing pseudo-asteroids that spacecraft can fly by in the same optimal trajectory. Finally, we apply beam search using the surrogate model to find good asteroid flyby sequences.


%
%
\subsection{Architecture of Earth-Asteroid-Earth Block}

\begin{figure}[htb]
\begin{center}
\includegraphics[width=0.5\hsize]{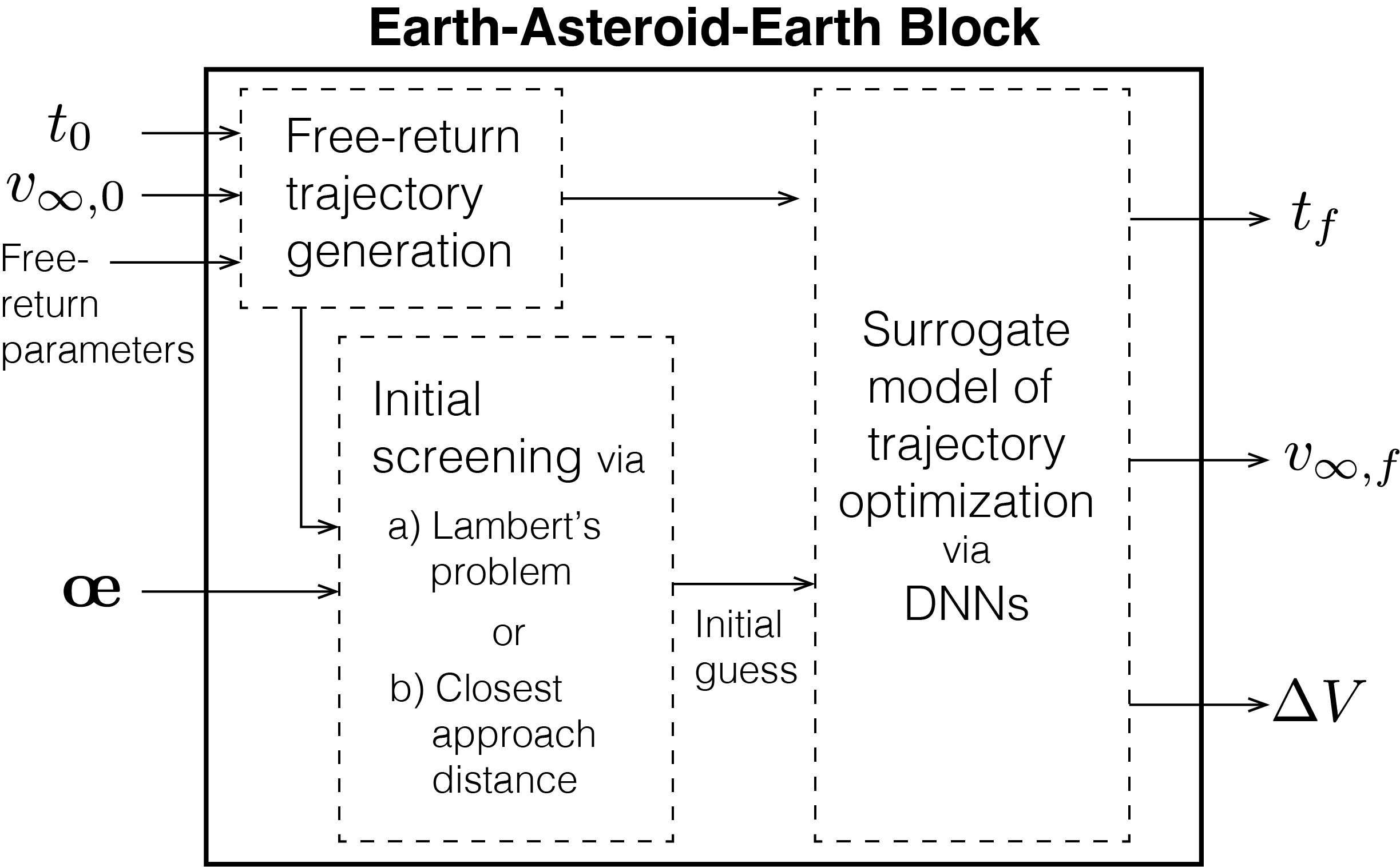}
\caption{\label{f:eae_block_architecture}Architecture of Earth-asteroid-Earth block.}
\end{center}
\end{figure}

In the Earth-asteroid-Earth block, we first generate the free-return trajectories for given parameters ($v_{\infty}$, the number of revolutions, the type classifiers of the free-return trajectories) defined in Table \ref{tab:freereturn_parameters}. The optimal Earth-asteroid-Earth trajectory exists near an Earth free-return trajectory. In particular, the free-return trajectory will be the optimal trajectory when the asteroid crosses in the free-return trajectory. Using the information of the free-return trajectory, we filter the accessible asteroids through screening algorithms based on a) Lambert's problem or b) closest approach distance from the free-return trajectory. We use the results of the screening algorithms as the initial guess of the trajectory optimization problem, that is, as the inputs of the surrogate model. We build a surrogate model of trajectory optimization because it is the most computationally intensive. Figure \ref{f:eae_block_architecture} illustrates the architecture of the Earth-asteroid-Earth block.

%
%
\subsection{Initial Screening Algorithm}

We adopt two initial screening algorithms and compare their performance. One method solves Lambert's problem, and the other calculates the closest approach distance from the free-return trajectory.

%
%
\subsubsection{Screening Algorithm by Lambert's Problem}

One of the asteroid screening algorithms that we adopt is based on Lambert's problem. We use the stable and fast computational method of Lambert's problem in Ref.\cite{Izzo2015}. This screening algorithm divides the Earth-asteroid-Earth trajectory into two phases; the first phase is from the Earth to the asteroid, and the second phase is from the asteroid to the Earth. We solve Lambert's problems for each phase by changing the asteroid flyby epoch. The initial and final epochs at the Earth flybys are fixed to the ones of the free-return trajectory. Using a grid search regarding the asteroid flyby epoch, we iteratively solve Lambert's problems to find the optimal flyby epoch, where the step size is 3 days in our numerical example. \majorred{We choose this step size as small as the computation time allows. Because the computation time of Lambert's problem is not negligible in the proposed architecture, we can likely accelerate the computation by introducing an adaptive step size scheme.} The objective function of this grid search is the sum of two $\Delta V$s needed for the transfer. One of the $\Delta V$s is the difference between the initial \majorblue{hyperbolic excess velocity} needed for Lambert's transfer and the one we assume for the free-return trajectory. The other $\Delta V$ is the velocity difference at the asteroid flyby point between phases. We store the results if the total $\Delta V$ is less than a threshold. The threshold is set to 3 km/s in our numerical example.

%
%
\subsubsection{Screening Algorithm by Closest Approach Distance}

The other asteroid screening algorithm that we adopt is based on the closest approach distance of the asteroid from an Earth free-return trajectory. In this algorithm, we first propagate a free-return trajectory and all asteroid trajectories. In this propagation, the initial epoch $t_0$ is given; the final epoch $t_{\textrm{FR}, f}$ is given from the free-return transfer; the time step $\Delta t$ is a tuned parameter, where $\Delta t$ is 3 days in our numerical example. For each fraction of the trajectories defined in the time span $(t_j, t_j+\Delta t)$, we calculate the distance $d(t_j)$ between the free-return trajectory and an asteroid trajectory under the uniform linear motion assumption by the following equation
\begin{align}
    d (t_j) &= \| \bm{r}_{\textrm{rel}} - \delta t\bm{v}_{\textrm{rel}}\|\\
    \delta t &\majorblue{:=}\min\left(\max\left(\frac{\bm{r}_{\textrm{rel}}\cdot \bm{v}_{\textrm{rel}}}{\|\bm{v}_{\textrm{rel}}\|^2}, 0\right), \Delta t\right)
\end{align}
where $\bm{r}_{\textrm{rel}}:=\bm{r}_{\textrm{sc}}(t_j) - \bm{r}_{\star}(t_j) $,  $\bm{v}_{\textrm{rel}}:=\bm{v}_{\textrm{sc}}(t_j) - \bm{v}_{\star}(t_j)$. Then, the closest approach distance $d_{\textrm{CA}}$ between the free-return trajectory and the asteroid is calculated by 
\begin{equation}
    d_{\textrm{CA}} = \min_{t_j\in\mathcal{T}} d(t_j)
\end{equation}
where $\mathcal{T}:=\left\{ t_0+ j\Delta t : j\in\mathbb{N}, t_0+j\Delta t \in(t_0, t_{\textrm{FR},f}) \right\}$. We store the results if the closest approach distance $d_{CA}$ is less than a threshold, which is 5,000,000 km in our numerical example.

%
%
\subsection{Generating Database of Earth-Asteroid-Earth Optimal Trajectories}

\begin{figure}[htb]
\begin{center}
\includegraphics[width=0.5\hsize]{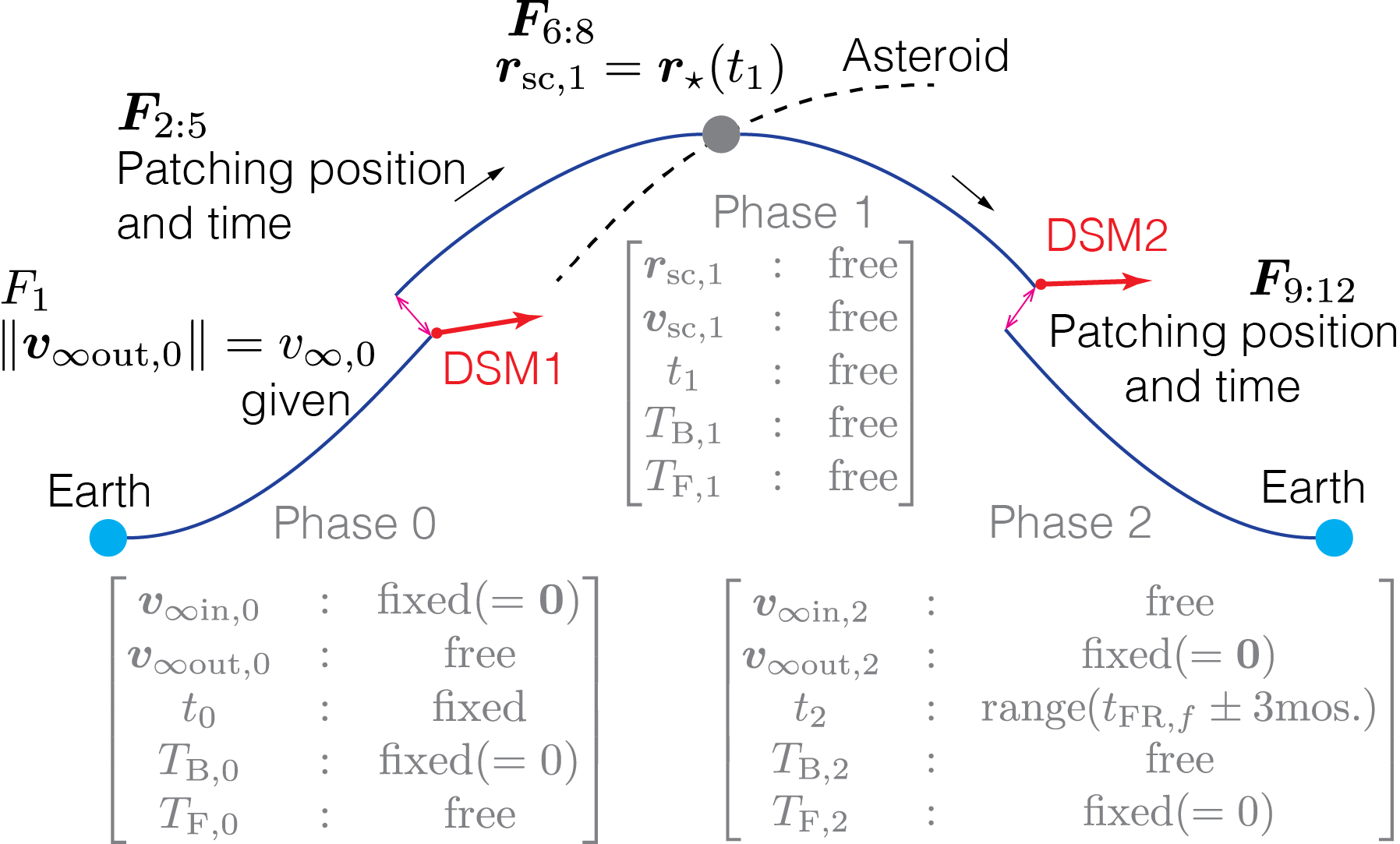}
\caption{\label{f:MGA1DSM_multipleshooting}Definition of trajectory optimization problem via multiple shooting.}
\end{center}
\end{figure}

We optimize the trajectory via a direct multiple shooting method using the initial guess trajectories produced by the initial screening algorithm. Lambert's solution gives the initial guess trajectory for Lambert's screening, while the free-return trajectory provides the initial guess trajectory for the closest approach screening. Figure \ref{f:MGA1DSM_multipleshooting} illustrates the definition of the trajectory optimization problem. The Earth-asteroid-Earth trajectory is divided into three phases. We introduce 9 optimization variables for each phase, including nodal state vectors \majorblue{$\left[\bm{r}_{\textrm{sc},i}^{\top}, \bm{v}_{\textrm{sc},i}^{\top}\right]^{\top}$} or \majorblue{$\left[\bm{v}_{\infty\textrm{in},i}^{\top}, \bm{v}_{\infty\textrm{out}, i}^{\top}\right]^{\top}$}, time $t_i$, backward propagation time \majorblue{$T_{\textrm{B},i}$}, and forward propagation time \majorblue{$T_{\textrm{F},i}$}. The objective function of this trajectory optimization is the sum of $\Delta V$s at the patching points
\begin{align}
    J &= \sum_{i=1}^2 \Delta V_i\nonumber\\
    &= \sum_{i=1}^2 \sqrt{\|\bm{v}_{\textrm{B},i} - \bm{v}_{\textrm{F},i-1} \|^2 + \epsilon},
\end{align}
where $\bm{v}_{\textrm{B},i}$ is the velocity calculated by the backward propagation of phase $i$;  $\bm{v}_{\textrm{F},i-1}$ is the velocity calculated by the forward propagation of phase $i-1$, and; $\epsilon$ is a small number introduced to make the objective function differentiable. 

We consider 12 equality constraints \majorblue{defined by the following equations.
\begin{align}
    F_1 &= \|\bm{v}_{\infty\textrm{out},0}\| - \bar{v}_{\infty,0} = 0\\
    F_2 &= \left(t_0 + T_{\textrm{F},0}\right) - \left(t_1 + T_{\textrm{B},1}\right) = 0\\
    \bm{F}_{3:5} &= \bm{r}_{\textrm{B},1} - \bm{r}_{\textrm{F},0} = \bm{0}\\
    \bm{F}_{6:8} &= \bm{r}_{\textrm{sc},1} - \bm{r}_{\star}(t_1) = \bm{0}\\
    F_9 &= \left(t_1 + T_{\textrm{F},1}\right) - \left(t_2 + T_{\textrm{B},2}\right) = 0\\
    \bm{F}_{10:12} &= \bm{r}_{\textrm{B},2} - \bm{r}_{\textrm{F},1} = \bm{0}
\end{align}
where $\bm{r}_{\textrm{B},i}$ is the position calculated by the backward propagation of phase $i$;  $\bm{r}_{\textrm{F},i-1}$ is the position calculated by the forward propagation of phase $i-1$; $\bar{v}_{\infty,0}$ is the v-infinity magnitude given as the input of the Earth-asteroid-Earth block, and; $\bm{r}_{\star}(t_1)$ is the position of the asteroid at $t_1$ calculated from its ephemeris. We bound the optimization variables as given by the following equations.
\begin{align}
    \bm{v}_{\infty\textrm{in},0} &= \bm{0}\\
    t_0 &= \bar{t}_0\\
    T_{\textrm{B},0} &= 0\\
    \bm{v}_{\infty\textrm{out},2} &= \bm{0}\\
    t_{\textrm{FR},f}-3\textrm{mos.} &\leq t_2 \leq t_{\textrm{FR},f} + 3\textrm{mos.}\\
    T_{\textrm{F},2} &= 0
\end{align}
where $\bar{t}_0$ is the initial epoch given as the input of the Earth-asteroid-Earth block, and; $t_{\textrm{FR},f}$ is the final epoch given from the free-return transfer.} Finally, the trajectory optimization problem can be described by the form of nonlinear programming (NLP), which can be solved by a sequential quadratic programming (SQP) solver such as SNOPT\cite{GilMS05}. 

\begin{figure}[htb]
\begin{center}
\includegraphics[width=0.45\hsize]{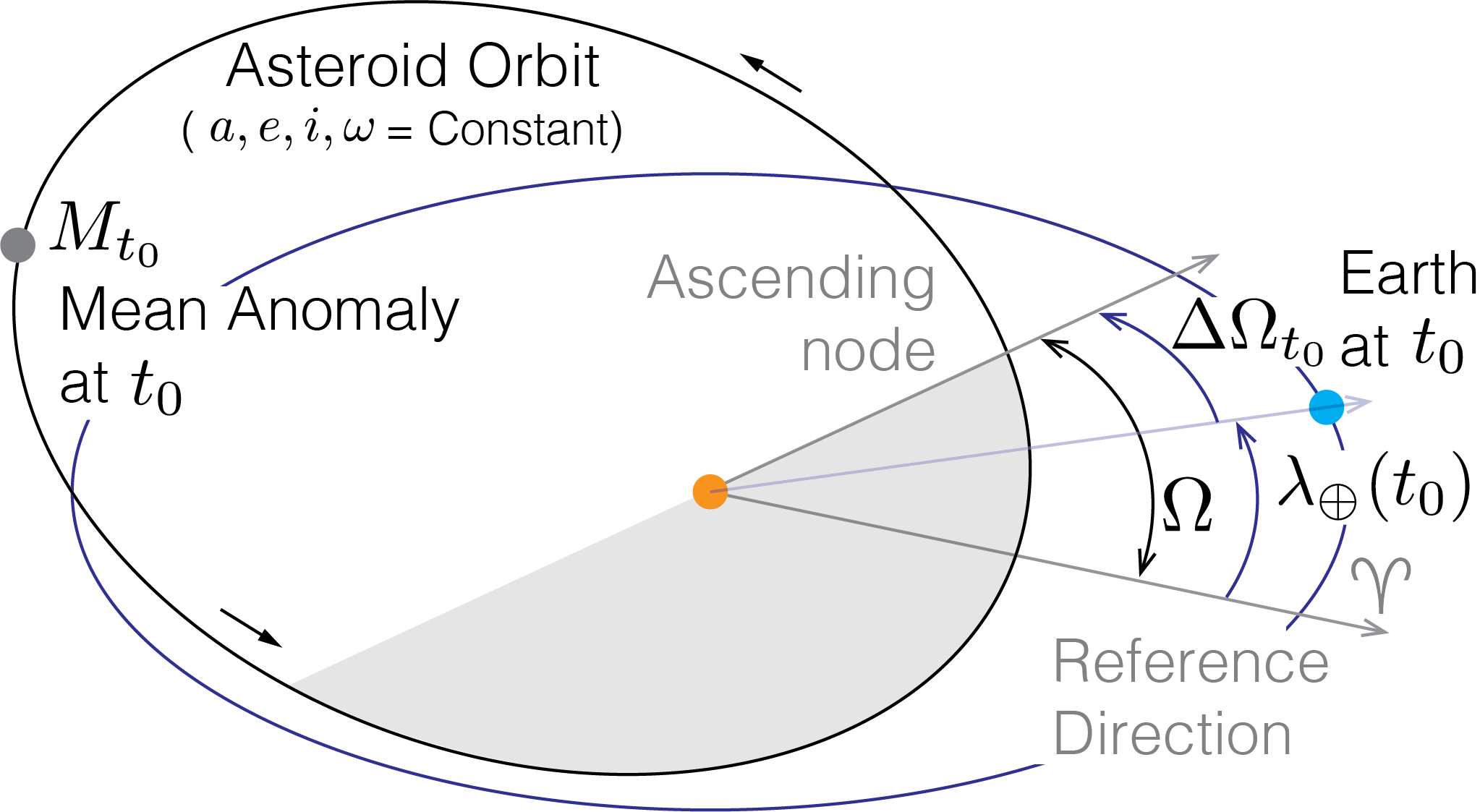}
\caption{\label{f:om_correction}Definition of $\Delta \Omega$ and mean anomaly at $t_0$.}
\end{center}
\end{figure}

The resulting database has a dependency on the initial epoch, i.e., the Earth departure epoch. When we use this database to build a surrogate model, the generalization performance will be insufficient if the Earth departure epoch is outside the range of the database. Therefore, as post-processing, we transform a database that is independent of the Earth departure epoch. As shown in Fig. \ref{f:om_correction}, the orbital elements $\Omega$ and $M$ of the asteroid are converted as follows.
\begin{equation}
    \begin{cases}
    \Omega \rightarrow \Delta \Omega_{t_0} := \Omega - \lambda_{\oplus} (t_0)\\
    M_{t_{\textrm{\oe}}} \rightarrow M_{t_0} := M_{t_{\textrm{\oe}}} + \sqrt{\frac{\majorblue{\mu}_{\odot}}{a^3}}(t_0-t_{\textrm{\oe}})
    \end{cases} \label{eq:omega_ma_corr}
\end{equation}
where \majorblue{$\lambda_{\oplus} (t_0)$ is the longitude of the Earth at $t_0$, and} $t_{\textrm{\oe}}$ is the epoch at which the orbital element is given. The remaining orbital elements $a, e, i, \omega$ are invariant to this transformation. \majorblue{This transformation allows the proposed surrogate model to be applied to any Earth departure epoch.}

%
%
\subsection{Extension of Database with Pseudo Asteroid under KKT conditions}


Since the database generation by trajectory optimization is computationally expensive, we propose a new database generation strategy by introducing pseudo-asteroids \majorblue{that can be flown by spacecraft} in the same optimal trajectory. The spacecraft can fly by a pseudo-asteroid if the pseudo-asteroid crosses the spacecraft's trajectory. However, this crossing condition does not guarantee the optimality of this trajectory to the pseudo-asteroid. In order to ensure the optimality of the trajectory, the pseudo-asteroid must satisfy the optimality conditions, also known as the \majorblue{KKT} conditions. The proposed approach generates pseudo-asteroids for each optimal trajectory, for which we have already computed all optimization variables and Lagrange multipliers.

The optimal trajectory satisfies the following constraint at the flyby point
\begin{equation}
    \bm{F}_{6:8} = \bm{r}_{\textrm{sc},1} - \bm{r}_{\star}(t_1) = \bm{0}\nonumber
\end{equation}
Let us assume that a pseudo-asteroid intersects the optimized trajectory at $t_1$. Then, the position of the pseudo-asteroid $\bm{r}_{\textrm{p}\star}$ can be computed as follows.
\begin{equation}
    \bm{r}_{\textrm{p}\star}(t_1) = \bm{r}_{\star}(t_1) = \bm{r}_{sc,1}
\end{equation}

The velocity of the pseudo-asteroid $\bm{v}_{\textrm{p}\star}$ is determined through the conservation of the KKT conditions. Because all optimization variables and Lagrange multipliers maintain the same value, the following terms of the KKT conditions must be satisfied.
\majorblue{
\begin{align}
    \bm{\lambda}_{6:8}^{\top}\frac{\partial \bm{F}_{6:8}}{\partial t_1} = \bm{\lambda}_{6:8}^{\top} \left\{-\bm{v}_{\star}(t_1)\right\}  = \bm{\lambda}_{6:8}^{\top} \left\{-\bm{v}_{\textrm{p}\star}(t_1)\right\}
\end{align}
}
Hence,
\majorblue{
\begin{align}
    \bm{\lambda}_{6:8}^{\top} \left\{\bm{v}_{\textrm{p}\star}(t_1)-\bm{v}_{\star}(t_1)\right\} = 0
\end{align}
}
Note that the velocity vector of the pseudo-asteroid $\bm{v}_{\textrm{p}\star}$ has two degrees of freedom. 

\begin{figure}[htb]
\begin{center}
\includegraphics[width=0.2\hsize]{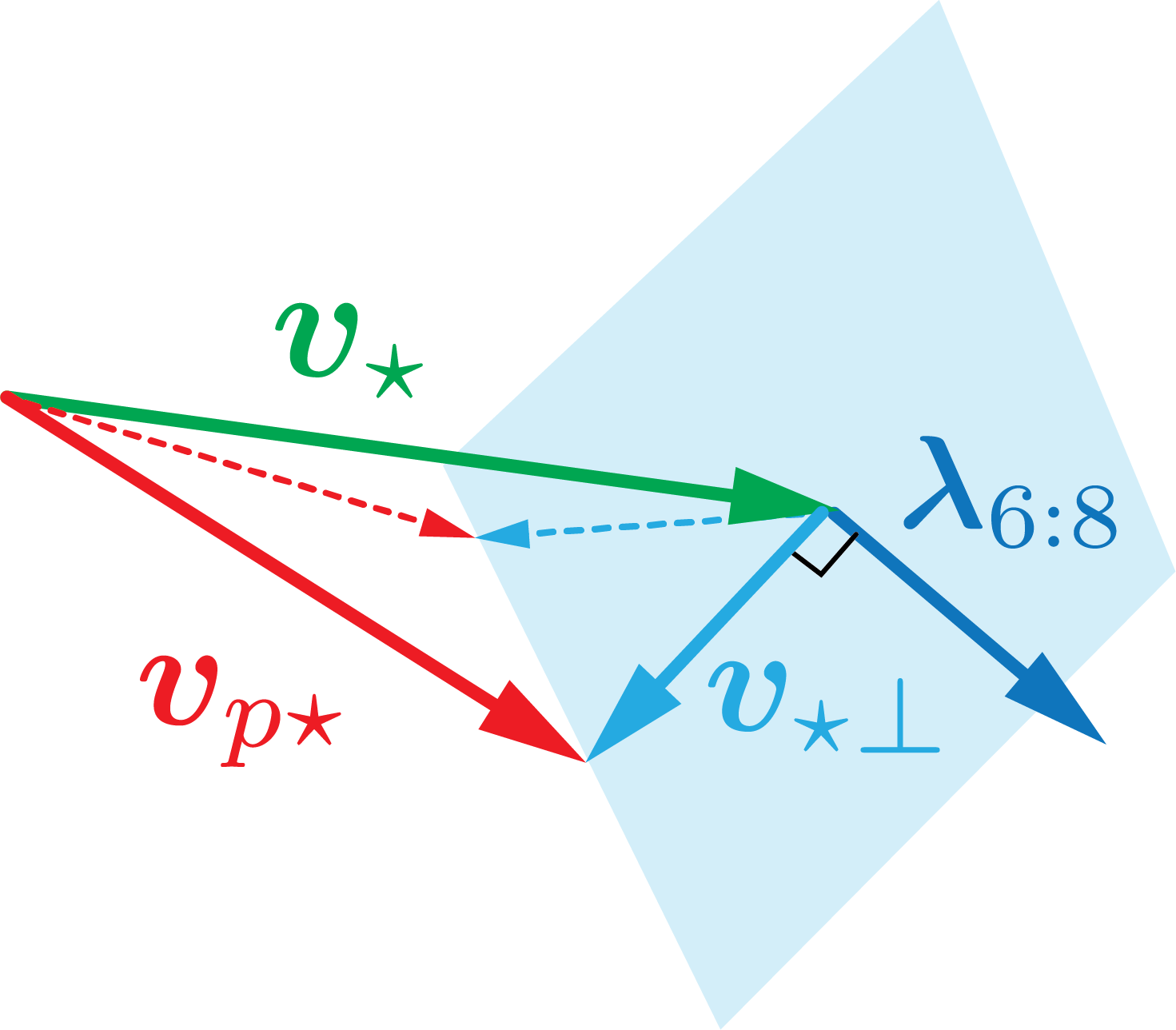}
\caption{\label{f:pseudo_asteroid}Pseudo-asteroid velocity.}
\end{center}
\end{figure}

We can determine the velocity of the pseudo-asteroids by introducing two random variables $\bm{\nu}_{\textrm{rand}}$ and $\alpha_{\textrm{rand}}$. The first step computes a vector perpendicular to $\bm{\lambda}_{6:8}$.
\majorblue{
\begin{align}
    \bm{v}_{\star\perp} = \bm{\nu}_{\textrm{rand}} - \frac{\bm{\lambda}_{6:8}^{\top}\bm{\nu}_{\textrm{rand}}}{\|\bm{\lambda}_{6:8}\|^2} \bm{\lambda}_{6:8}
\end{align}
}
The second step calculates the velocity of the pseudo-asteroid as the following equation.
\begin{equation}
    \bm{v}_{\textrm{p}\star}(t_1) = \bm{v}_{\star}(t_1) + \alpha_{\textrm{rand}} \bm{v}_{\star\perp}
\end{equation}
The relation among $\bm{v}_{\star}, \bm{v}_{\textrm{p}\star}, \lambda_{6:8}$, and $\bm{v}_{\star\perp}$ is illustrated in Fig. \ref{f:pseudo_asteroid}. In the numerical results, we set up the bound on the semi-major axis of the pseudo-asteroid. Therefore, $\alpha_{\textrm{rand}}$ is decided so that $\bm{v}_{\textrm{p}\star}$ satisfies
\begin{equation}
    a_{\textrm{LB}} \leq \left(\frac{2}{\|\bm{r}_{\textrm{p}\star}\|}-\frac{\|\bm{v}_{\textrm{p}\star}\|^2}{\majorblue{\mu}_{\odot}}\right)^{-1} \leq a_{\textrm{UB}}.
\end{equation}

We can finally calculate the Keplerian orbital elements of the pseudo-asteroid $\textbf{\oe}_{\textrm{p}}$ from the Cartesian state vector $(\bm{r}_{\textrm{p}\star}, \bm{v}_{\textrm{p}\star})$ at the epoch $t_1$. If we generate 10 pseudo-asteroids for each optimal trajectory, the proposed method can expand the size of the database about 11 times.

%
%
\subsection{Surrogate Model by Deep Neural Networks}\label{sec:method_dnn}

We create the surrogate model by the DNNs regression using the massive databases of the optimal trajectories. \majorred{In this work we use a feed-forward neural network with fully-connected layers, and hyperparameters such as the number of layers $M$, the number of units $N$, and the learning rate $\epsilon$ are selected through the sensitivity analysis.} Activation functions of the first $(M-1)$ layers are exponential linear units (ELUs)\cite{clevert2015fast} and of the last one is sigmoid to handle the bounded outputs. We add an dropout or a one-dimensional batch normalization \cite{ioffe2015batch} layer after each of the first \majorblue{$(M-1)$ layers}. The dropout layer is to prevent over-fitting and the batch normalization layer is to improve stability.

The inputs and outputs of the regression model are shown in Tables \ref{tab:lambert_dnn} and \ref{tab:distance_dnn}. The details of the input information are explained in Appendix A. We apply two operations to the outputs in order to improve the performance of the regression. The first operation calculates the difference of $t_f$ and $v_{\infty,f}$ from the values of the free-return trajectories. That is,
\begin{align}
    \begin{cases}
        \Delta t_f &= t_f - t_{f\textrm{FR}}\\
        \Delta v_{\infty,f} &= v_{\infty, f} - v_{\infty, f\textrm{FR}}.
    \end{cases}
\end{align}
where $t_{f\textrm{FR}}$ and $v_{\infty, f\textrm{FR}}$ are respectively the final epoch and \majorblue{hyperbolic excess velocity} of the free-return trajectory calculated from $(m, n, \textrm{type}, v_{\infty,0})$. Note that we do not apply this operation to $\Delta V$ because $\Delta V$ of the free-return trajectory is zero. This operation lets \majorblue{$\Delta \bm{z} \left(= [\Delta t_f, \Delta v_{\infty, f}, \Delta V]^{\top}\right)$} be distributed around zero. The second operation takes $\tan^{-1}(\cdot)$ for each element of $\Delta \bm{z}$. $\tan^{-1}(\cdot)$ provides a soft bound on the output and increases the sensitivity to values close to zero. Although $\Delta \bm{z}$ are unbound, the ranges of interest are fixed. For example, a trajectory design with $\Delta V =$10 km/s (out of the range of interest) usually does not require an accurate $\Delta V$ prediction. Finally, the outputs $\mathcal{Y}_{z_i}$ of DNNs are calculated as
\begin{equation}
    \mathcal{Y}_{z_i} = \tan^{-1}\left(\Delta z_i/\chi_{z_i}\right)
\end{equation}
where
\begin{equation}
    \chi_{z_i} = \Delta z_{i,90\%} \cdot \tan\left(\frac{0.9 \pi}{2}\right)
\end{equation}
where $\Delta z_i$ is an element of $\Delta \bm{z}$, and $\Delta z_{i,90\%}$ is an user-defined parameter.

For the stability of the training, input variables except for $m$, $n$, and "type" are normalized to a standard normal distribution and all output variables are re-scaled to $[0, 1]$.
About exceptions, $m$ and $n$ are used as is, and "type" is converted to an integer.

\begin{table}[hbt!]
\caption{\label{tab:lambert_dnn} Inputs and outputs of DNNs (screened by Lambert's problem)}
\centering
\begin{tabular}{ll}
\hline
Inputs & Outputs \\ \hline
Free-return info ($m, n$, type, $v_{\infty,0}$) & $\tan^{-1}\left(\Delta t_f/\chi_{\Delta t_f}\right)$\\
Asteroid ephemeris ($a, e, i, \Delta\Omega_{t_0}, \omega, M_{t_0}$) & $\tan^{-1}\left(\Delta v_{\infty, f}/\chi_{\Delta v_{\infty, f}}\right)$\\
$\Delta V$s of Lambert's Method ($\Delta v_0$, $\Delta v_1$, $\Delta v_{\textrm{total}}$) & $\tan^{-1}\left(\Delta V/\chi_{\Delta V}\right)$\\
Initial guess of $\majorblue{T}, n_{\star}, \eta_{t_{\star}}, \bm{v}_{\infty\textrm{out},0},  \bm{v}_{\infty\textrm{in},f}, \bm{v}_{\textrm{rel,in},1}$, and $\bm{v}_{\textrm{rel,out},1}$ & \\
\hline
\end{tabular}
\end{table}

\begin{table}[hbt!]
\caption{\label{tab:distance_dnn} Inputs and outputs of DNNs (screened by closest approach distance)}
\centering
\begin{tabular}{ll}
\hline
Inputs & Outputs \\ \hline
Free-return info ($m, n$, type, $v_{\infty,0}$) & $\tan^{-1}\left(\Delta t_f/\chi_{\Delta t_f}\right)$\\
Asteroid ephemeris ($a, e, i, \Delta\Omega_{t_0}, \omega, M_{t_0}$) & $\tan^{-1}\left(\Delta v_{\infty, f}/\chi_{\Delta v_{\infty, f}}\right)$ \\
Closest approach state difference ($\delta \bm{r}_{\textrm{CA}}$ and $\delta \bm{v}_{\textrm{CA}}$) & $\tan^{-1}\left(\Delta V/\chi_{\Delta V}\right)$\\
Initial guess of $\majorblue{T}, n_{\star}, \eta_{t_{\star}}, \bm{v}_{\infty,0},  \bm{v}_{\infty,f}$, and $\textbf{\oe}_{\textrm{FR}}$ &  \\
\hline
\end{tabular}
\end{table}

For training, we divide the entire dataset into minibatches \majorblue{, where the minibatch size is 1024,} and update the network parameters for each minibatch. The performance of the networks is evaluated by the mean squared error (MSE) loss defined as follows.
\begin{align}
    \mathcal{L}_{\textrm{total}} &:= \mathcal{L}_{\Delta t_f} +  \mathcal{L}_{\Delta v_{\infty,f}} +  \mathcal{L}_{\Delta V}\\
    \mathcal{L}_{\Delta t_f} &:= \frac{1}{3}\Braket{\left(\mathcal{Y}_{\Delta t_f}-\hat{\mathcal{Y}}_{\Delta t_f}\right)^2}\\
    \mathcal{L}_{\Delta v_{\infty,f}} &:= \frac{1}{3}\Braket{\left(\mathcal{Y}_{\Delta v_{\infty,f}}-\hat{\mathcal{Y}}_{\Delta v_{\infty,f}}\right)^2}\\
    \mathcal{L}_{\Delta V} &:= \frac{1}{3}\Braket{\left(\mathcal{Y}_{\Delta V}-\hat{\mathcal{Y}}_{\Delta V}\right)^2}
\end{align}
where $\hat{\cdot}$ is the estimated value by DNNs and we use the notation $\Braket{\cdot} = (1/N)\sum(\cdot)$ to describe the mean value across the minibatch. The network parameters are optimized to minimize the loss function using Adam\cite{kingma2014adam}, a stochastic gradient descent method. We run this training process over multiple epochs, where one epoch is completed when the entire dataset is consumed.




%
%
\subsection{Tree Search Method}

\begin{figure}[htb]
\begin{center}
\includegraphics[width=0.5\hsize]{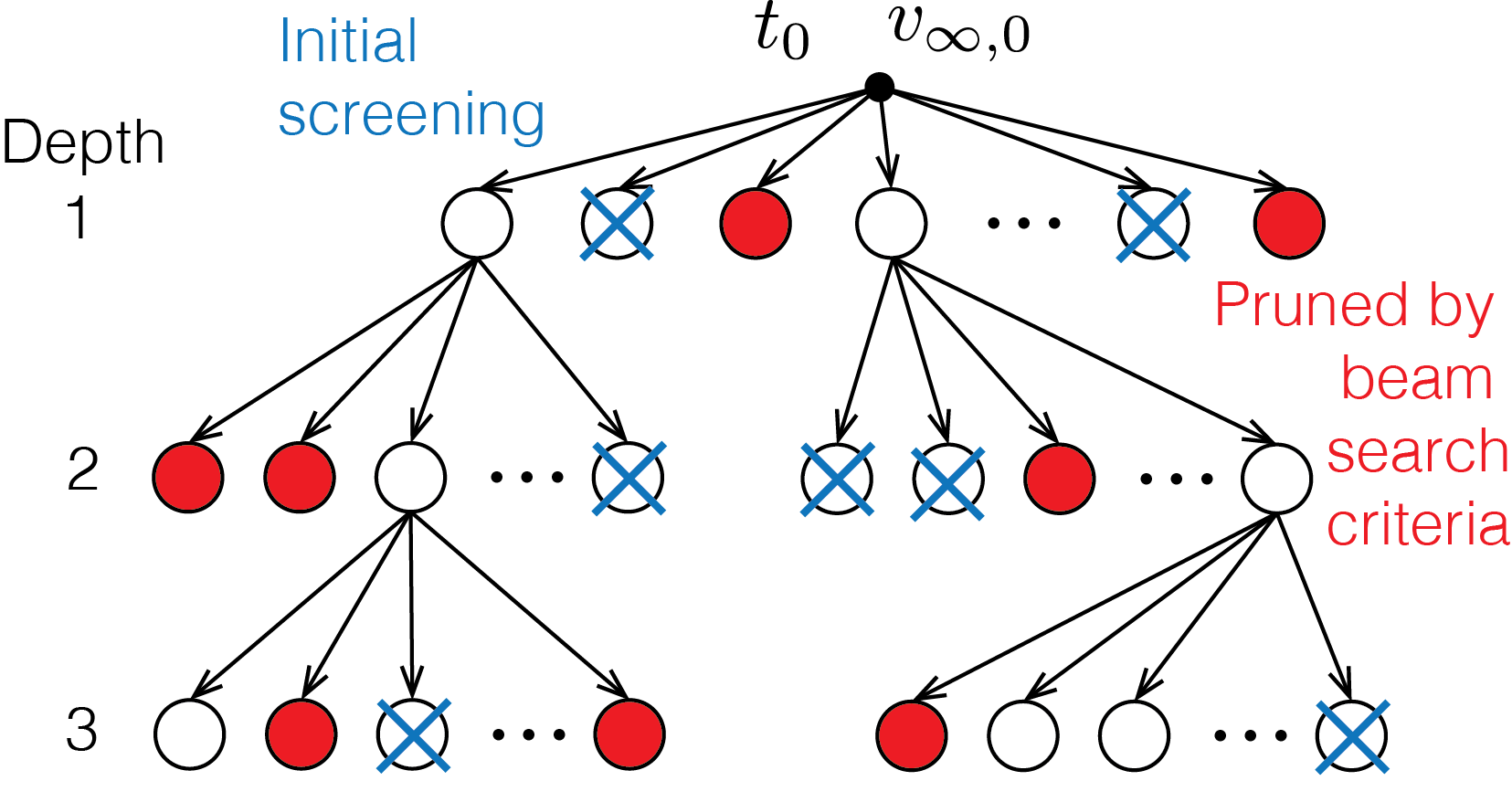}
\caption{\label{f:beam_search}Beam search method.}
\end{center}
\end{figure}

This paper employs beam search \majorblue{(e.g., \cite{Izzo2016})} to search for the good sequences of asteroids using the \majorblue{surrogate-based} Earth-asteroid-Earth block. Beam search is a heuristic search algorithm that uses breadth-first search in a limited set. At each level of the tree, the algorithm sorts successors by a heuristic cost, the total $\Delta V$ \majorblue{evaluated by the surrogate model}, and stores only a predetermined number of best states, called the {\it beam width}. The algorithm also discards the states if the total Time of Flight (TOF) is more than the upper bound (10 years in the numerical example) or the deflection angle of the Earth gravity assist is infeasible (minimum perigee altitude is 500 km in the numerical example). Beam search does not guarantee to find the best solution; however, it is helpful to systematically search the solutions. Figure \ref{f:beam_search} summarizes the beam search algorithm search for the asteroid flyby sequences. The initial screening algorithm first prunes unpromising solutions before evaluating the heuristic cost.

%
%

\section{Numerical Results}

\begin{figure}[htb]
\begin{center}
\includegraphics[width=0.45\hsize]{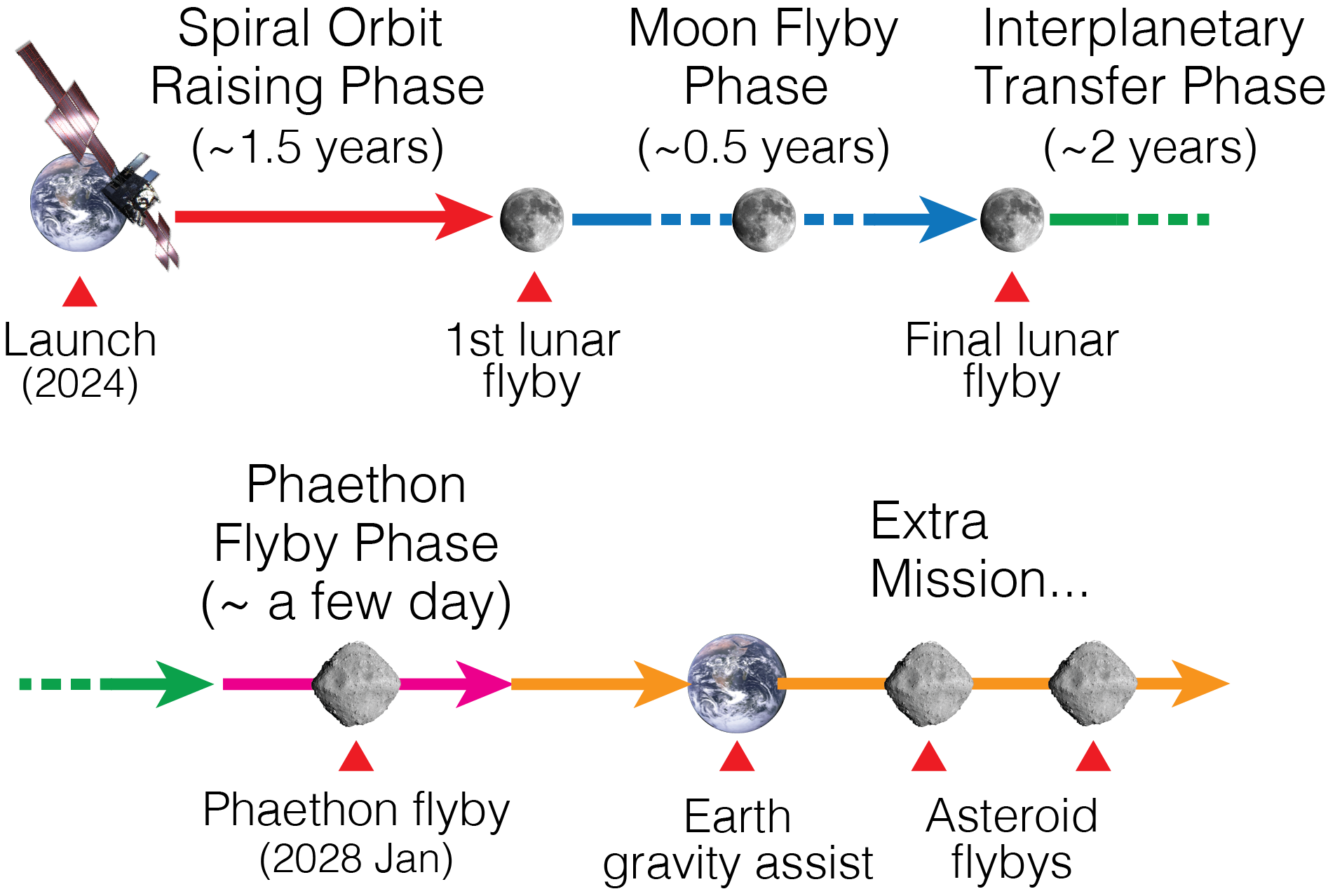}
\caption{\label{f:mission_scenario}Mission scenario of \destiny.}
\end{center}
\end{figure}

\begin{figure}
\subfigure[Near Earth trajectory (Earth-centered, ECLIPJ2000)]{%
    \includegraphics[clip, width=0.5\columnwidth]{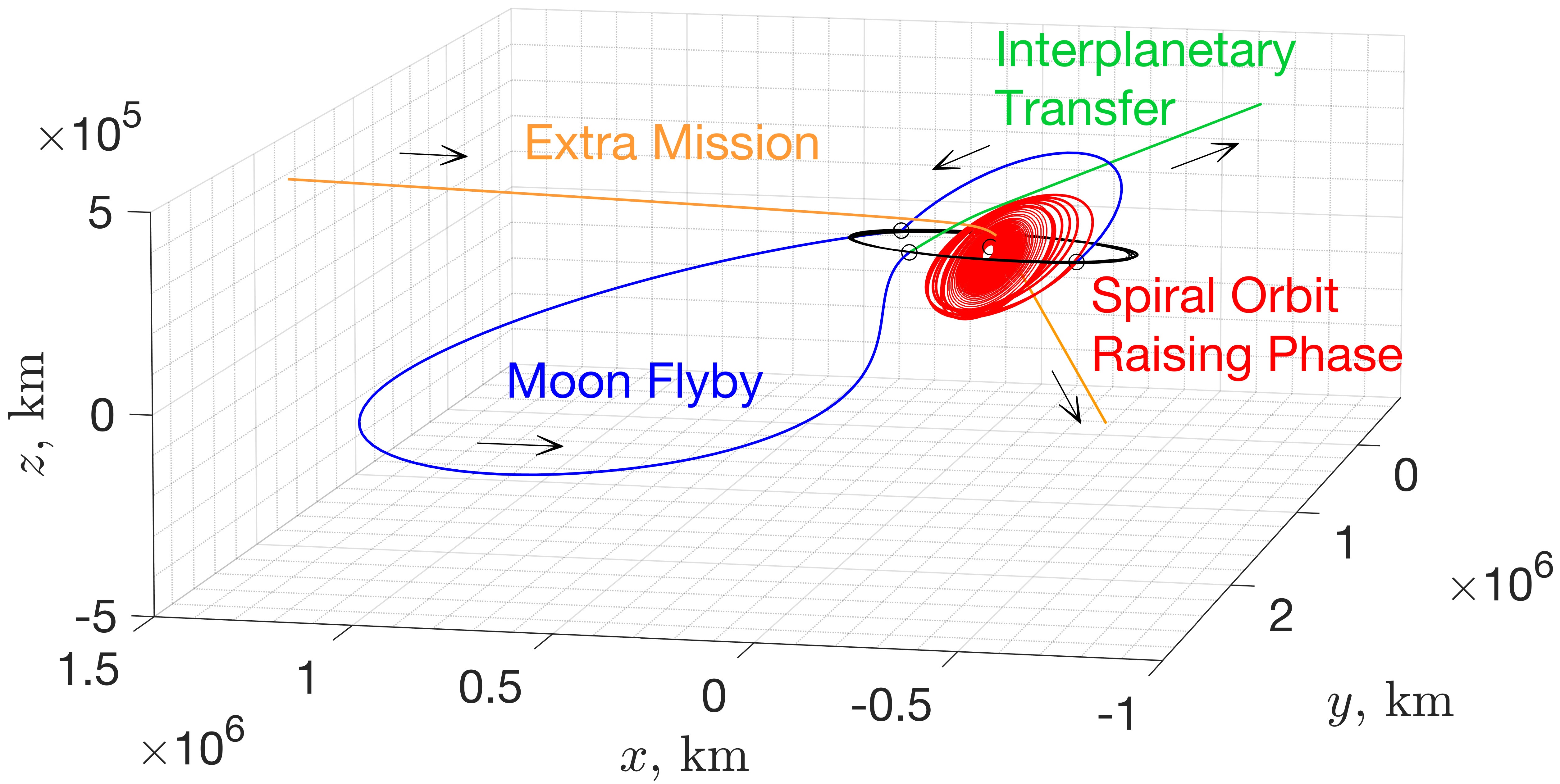}}%
\subfigure[Interplanetary trajectory (Sun-centered, ECLIPJ2000)]{%
    \includegraphics[clip, width=0.5\columnwidth]{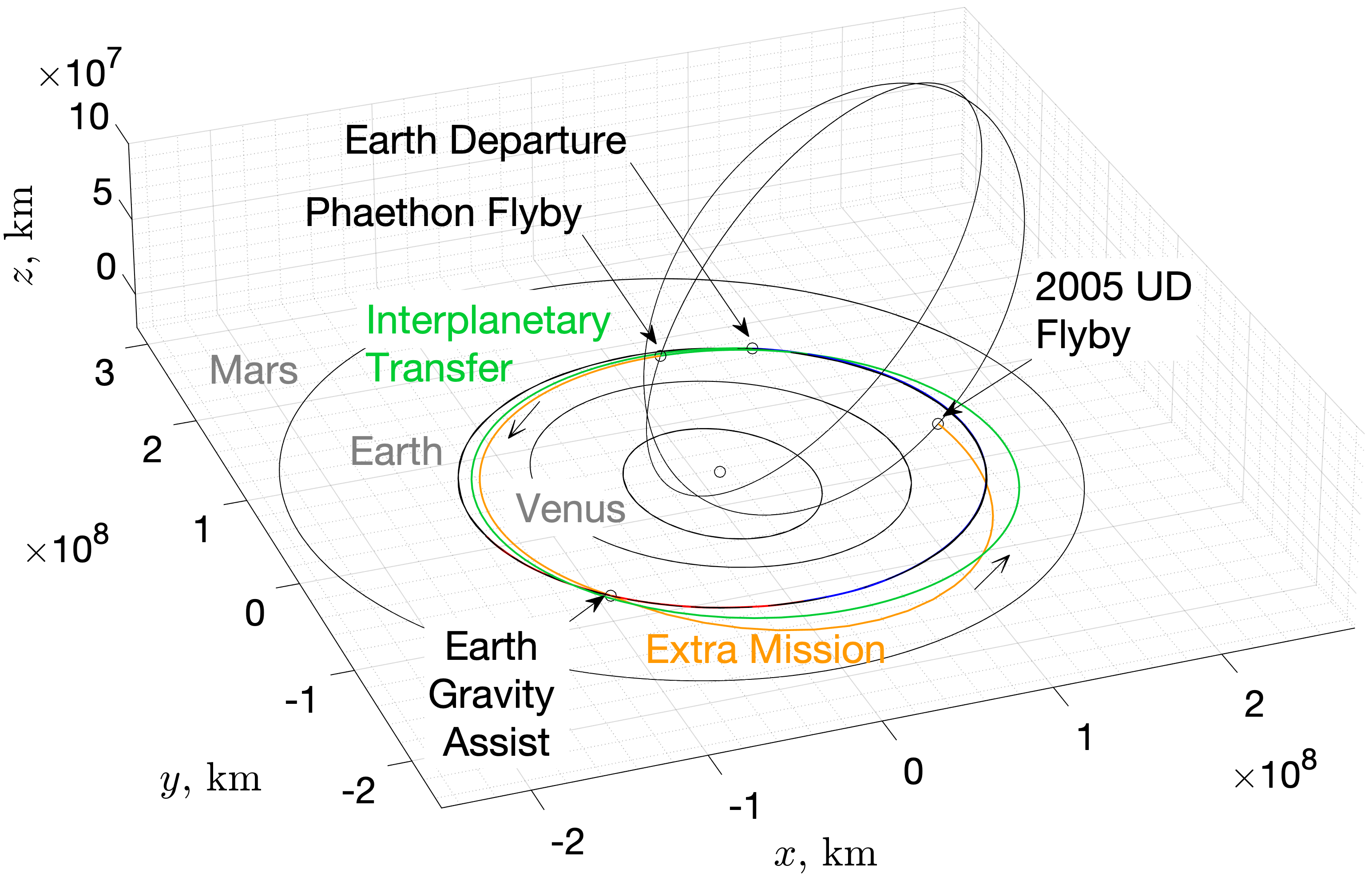}}%
\caption{\destiny baseline trajectory}
\label{f:dp_baseline_trajectory}
\end{figure}

This section applies the proposed method to \destiny mission\cite{Ozaki2022}, which is JAXA's Epsilon \majorblue{medium}-class mission to be launched in 2024. To enable lower cost and higher frequency deep space missions, the spacecraft will demonstrate advanced technologies that include highly efficient solar electric propulsion. For the science mission, the spacecraft will perform high-speed flyby observation and explore the asteroid (3200) Phaethon as the nominal mission and several more asteroids as an extra mission\cite{Ozaki2022, Sarli2015, Celik2021}. Figures \ref{f:mission_scenario} and \ref{f:dp_baseline_trajectory} respectively show the mission scenario and the baseline trajectories of \destiny. In this numerical example, we deal with the extra mission phase that starts with a gravity assist from the Earth after the Phaethon flyby.

%
%
\subsection{Precondition}


\destiny can only approach Phaethon at the vicinity of its descent node on the ecliptic plane because Phaethon's orbit has a large inclination angle and high eccentricity ($i=22.26$deg and $e=0.88990$). Considering the constraint on the distance to Earth for high-speed communication, the possible flyby timing is limited to either January 2028 (nominal) or November 2030 (backup). In the nominal scenario, the spacecraft returns to the Earth at 2028 MAY 05 12:13:59 TDB with $v_{\infty,0} = 2.684$km/s. We perform the following numerical simulations under this initial condition.

%
%
\subsection{Training Database}

We created multiple databases of the Earth-asteroid-Earth optimal trajectories (with each screening algorithm, with or without the pseudo-asteroids) and compared the performance of the surrogate models trained on each database. Table \ref{tab:cases_database} summarizes the definition of the database. \majorblue{Cases 2, 3, 5, 6, and 7 introduce pseudo-asteroids to efficiently increase the size of the database. Once we find the optimal trajectory, we generate 10 pseudo-asteroids that spacecraft can fly by in the same trajectory. By introducing 10 pseudo-asteroids per each trajectory optimization, we can multiple the size of the database by $(1+10)$ with low computation time.} For reference, Table \ref{tab:cases_database} also shows the data generation speed using the parallel computation on a 3.0 GHz Intel Core i9 workstation with 18 cores/36 threads. Introducing pseudo-asteroids speeds up the computational time significantly and therefore increases the database size greatly.

\begin{table}[hbt!]
\caption{\label{tab:cases_database} Definition of database}
\centering
\begin{tabular}{lcccc}
\hline
Case \# & Screening algorithm & pseudo-asteroids & Size of database & Speed, sample/s \\\hline
1 & Lambert's problem & w/o pseudo-asteroids & 698,368 & 8.77 \\ 
2 & Lambert's problem & w/ \majorblue{10} pseudo-asteroids & 698,368 & 47.6 \\ 
3 & Lambert's problem & w/ \majorblue{10} pseudo-asteroids & 6,983,680 & 47.6 \\ 
4 & Closest approach distance &  w/o pseudo-asteroids & 698,368 & 5.75 \\ 
5 & Closest approach distance & w/ \majorblue{10} pseudo-asteroids & 698,368 & 36.5 \\ 
6 & Closest approach distance & w/ \majorblue{10} pseudo-asteroids & 6,983,680 & 36.5 \\ 
7 & Lambert's problem & w/ \majorblue{10} pseudo-asteroids & 11,730,489 & 47.6 \\ 
\hline
\end{tabular}
\end{table}

As the post-process of the database generation, we remove outliers which do not satisfy the condition $0.5<a<10$au and $\Delta V<$10km/s, and apply the $\tan^{-1}(\cdot)$ operation to the outputs with $\Delta t_{f,90\%}=30$ days, $\Delta v_{\infty,f,90\%}=2$ km/s, and $\Delta V_{90\%}=1$ km/s.

\begin{figure}
\subfigure[Case 1: Lambert's problem w/o pseudo-asteroids]{%
    \includegraphics[clip, width=0.5\columnwidth]{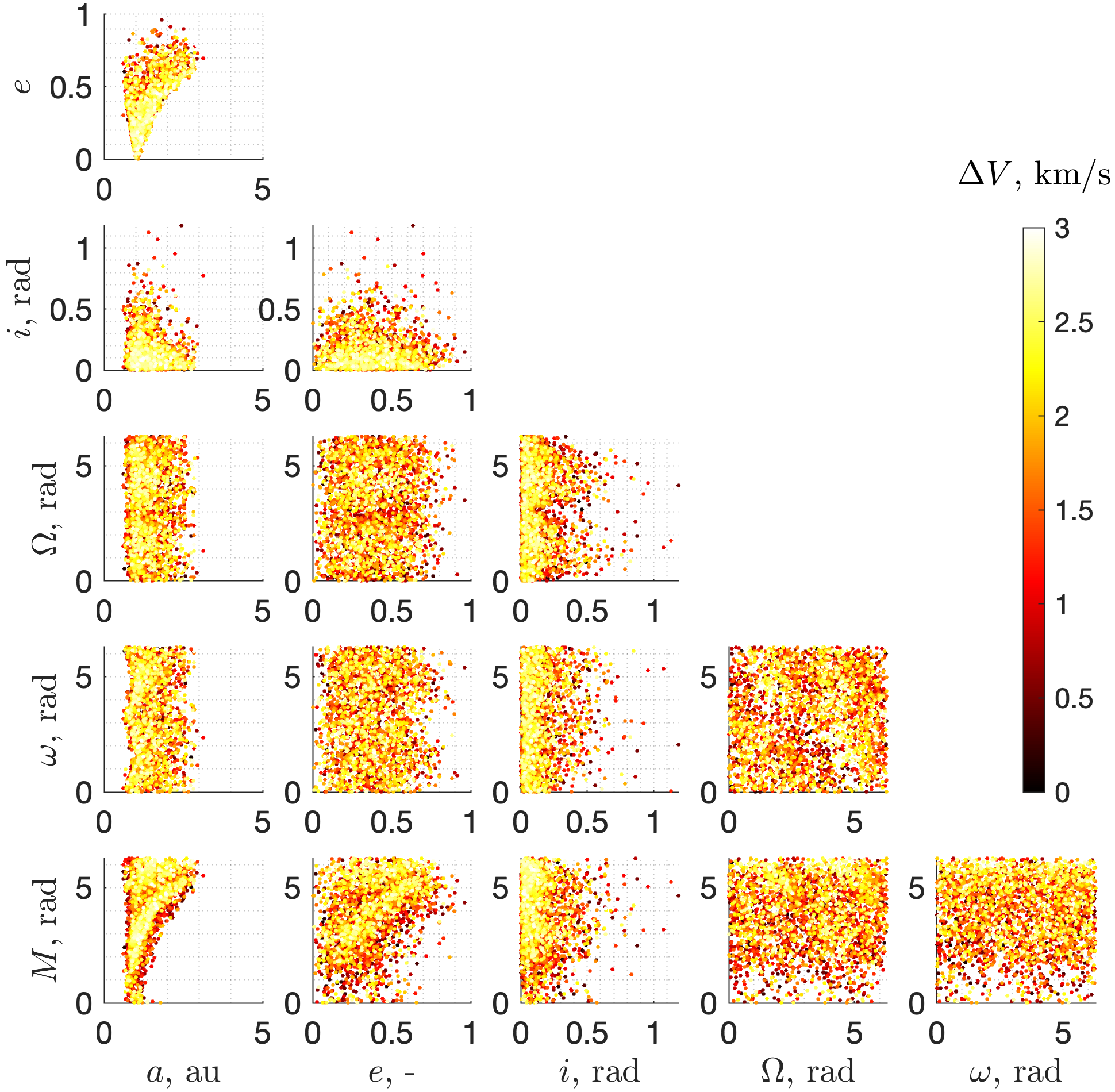}}%
\subfigure[Case 7: Lambert's problem w/ pseudo-asteroids]{%
    \includegraphics[clip, width=0.5\columnwidth]{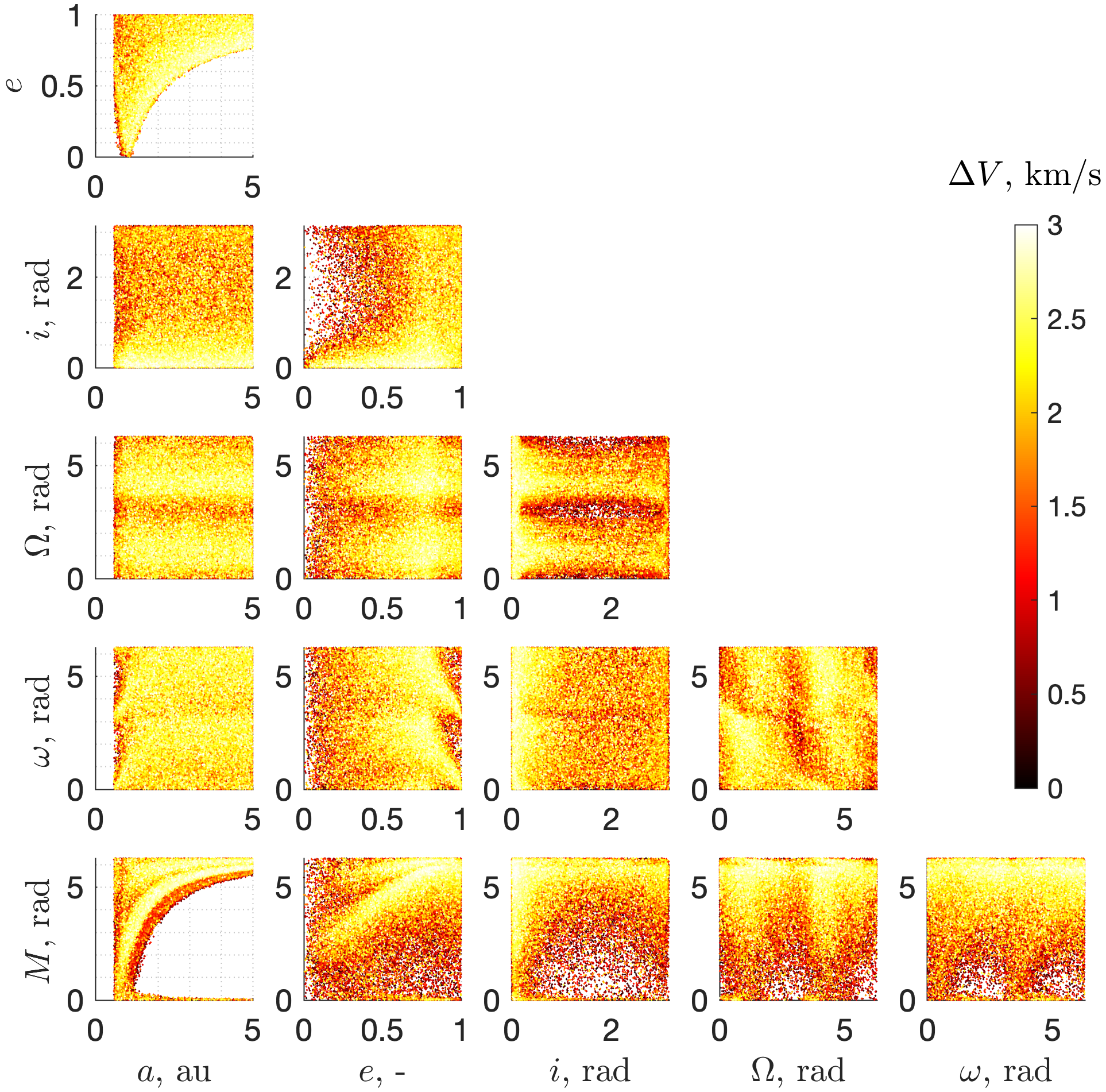}}%
\caption{$\Delta V$ plots of training databases (1:1 generic free-return trajectories).}
\label{f:training_data_plots}
\end{figure}

Figure \ref{f:training_data_plots} visualizes the databases for Case 1 and Case 7 with $\Delta V$ mapped to the asteroid orbital elements. The proposed method using pseudo-asteroids not only increases the size of the database, but also reduces the sparsity of the database that naturally exists in asteroid orbital dynamics.


%
%
\subsection{Performance of Neural Network}

\begin{figure}[htb]
\begin{center}
\includegraphics[width=0.5\hsize]{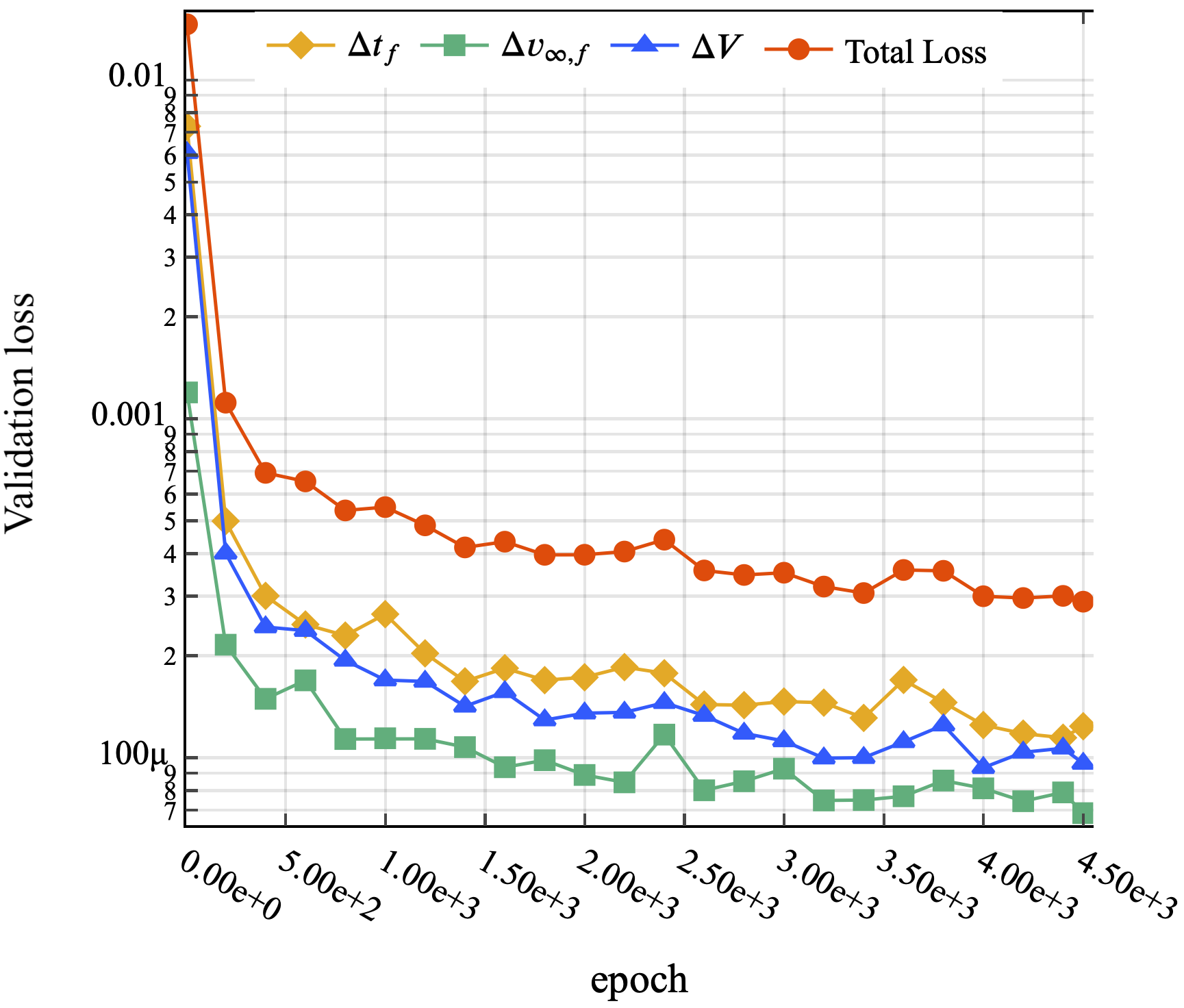}
\caption{\label{f:learning_curve_best}Learning curve of best case (case 7).}
\end{center}
\end{figure}

This subsection shows the results of training DNNs on the database shown in the previous subsection. We use 90\% of the databases as training data and 10\% of them as validation data. \majorred{Using a preliminary dataset independent of the datasets defined in Table \ref{tab:cases_database}, we first manually tune the hyperparameters of the networks and set the learning rate to 1e-4, the number of layers to 5, and the number of units to 1,024 for all cases.} \majorblue{The summary of the DNN performance is shown in Table \ref{tab:cases_dnn_training}, and the learning curves are illustrated in Appendix B.} \majorred{For the datasets of Cases 1, 2, 4, 5, DNNs with the batch normalization layer showed overfitting behavior, in which the validation loss deteriorated by more than one order of magnitude relative to the training loss. For these cases, introducing the dropout layer suppressed the overfitting behavior. On the other hand, applying the dropout layer worsened the validation loss compared to the batch normalization case. When the dataset size is larger than 7e6, the overfitting does not occur even with the batch normalization. Thus, we apply the dropout layer for Cases 1, 2, 4, 5 and the batch normalization layer for Cases 3, 6, 7. This result demonstrates that at least 7e6 datasets are needed to obtain practical performance for our application.} Comparing Cases 1, 2, 4, 5, we observe that the performance is slightly better when the pseudo-asteroids are introduced, even if the size of the database is the same. Including the larger databases, Cases 3 and 6, we find that Case 3 records the best performance. Since Case 3 is the best condition, we create a more extensive database Case 7 with the same conditions as Case 3 except the size of the database. The learning curves of Case 7 are shown in Fig.\ref{f:learning_curve_best}. Note that the total loss of Case 7 becomes 3.28e-4 at epoch 4.5k. For reference, the computational speed is about 200,000 samples/s for each epoch using the GPU computation on a 3.0 GHz Intel Core i9 workstation with NVIDIA Quadro GV100 (32GB HBM2, Tensor 118.5Tflops). \majorred{For Case 7, we performed an additional sensitivity analysis of the hyperparameters shown in Appendix C. Although we find better hyperparameters that reduce the validation loss, we perform the final evaluation using the parameters we initially set, considering validation loss and calculation speed of training and predicting.} 



\begin{table}[hbt!]
\caption{\label{tab:cases_dnn_training} Performance of deep neural networks}
\centering
\begin{tabular}{lccc}
\hline
Case \# & Size of database & BN/DO & Validation loss (@epoch) \\\hline
1 & 698,368 & Drop Out & 2.26e-2 (@1k) \\ 
2 & 698,368 & Drop Out & 1.60e-2 (@1k) \\ 
3 & 6,983,680 & Batch Norm & 6.04e-4 (@1k) \\ 
4 & 698,368 & Drop Out & 1.70e-2 (@1k) \\ 
5 & 698,368 & Drop Out & 1.41e-2 (@1k) \\ 
6 & 6,983,680 & Batch Norm & 9.36e-3 (@1k) \\ 
7 & 11,730,489 & Batch Norm & 5.49e-4 (@1k), 3.28e-4(@4.5k) \\ 
\hline
\end{tabular}
\end{table}

\begin{figure}
\subfigure[Estimated $\Delta V$ by DNN vs true optimal $\Delta V$]{%
    \includegraphics[clip, width=0.45\columnwidth]{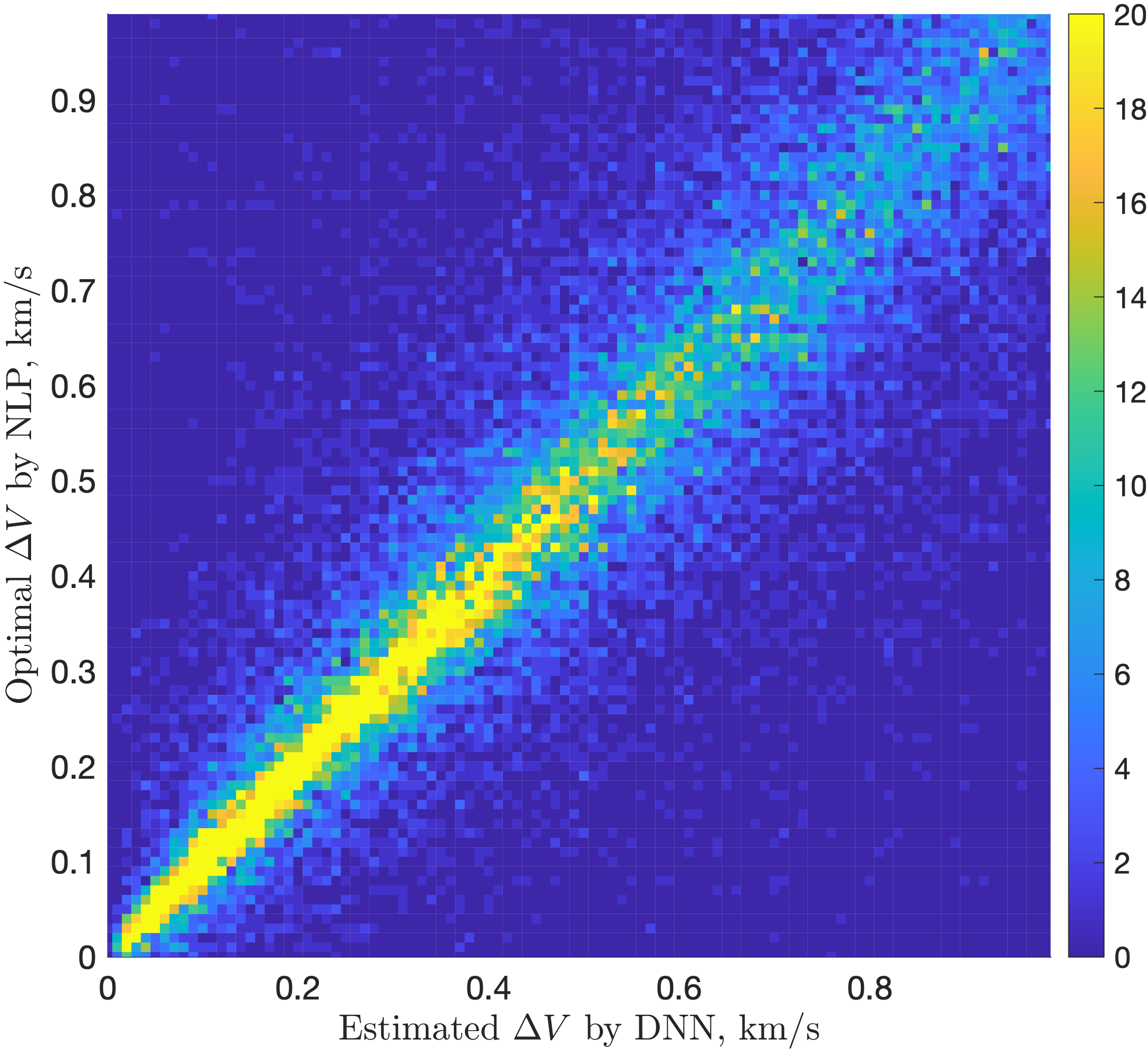}}%
\subfigure[Estimated $\Delta V$ by Lambert's problem vs true optimal $\Delta V$]{%
    \includegraphics[clip, width=0.45\columnwidth]{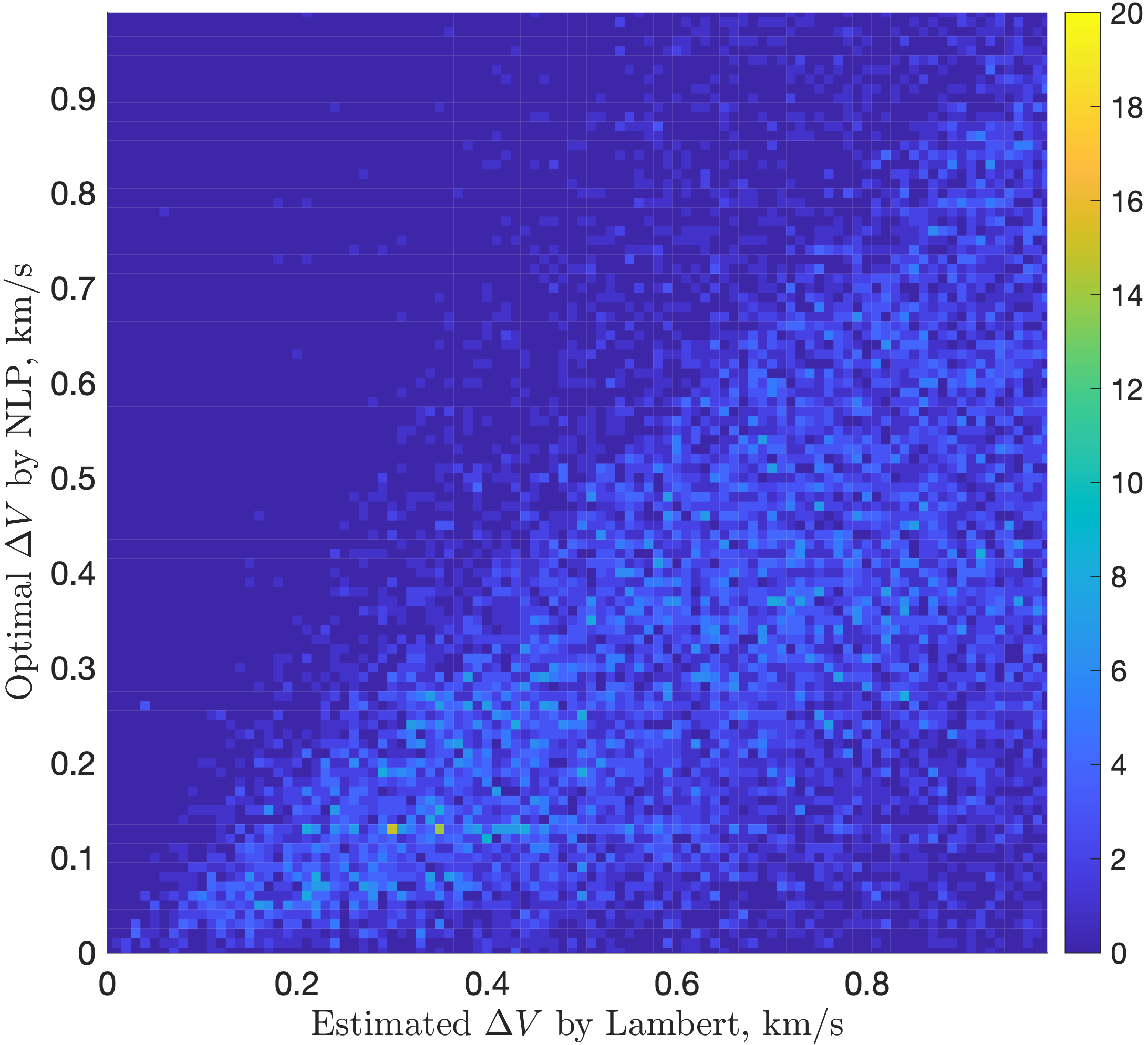}}%
\caption{\majorblue{Density heatmaps} of the estimated $\Delta V$s vs the true $\Delta V$s (Color map: counts of the solution).}
\label{f:dv_compare}
\end{figure}

\begin{figure}
\subfigure[Estimated $\Delta t_f$ vs true $\Delta t_f$]{%
    \includegraphics[clip, width=0.45\columnwidth]{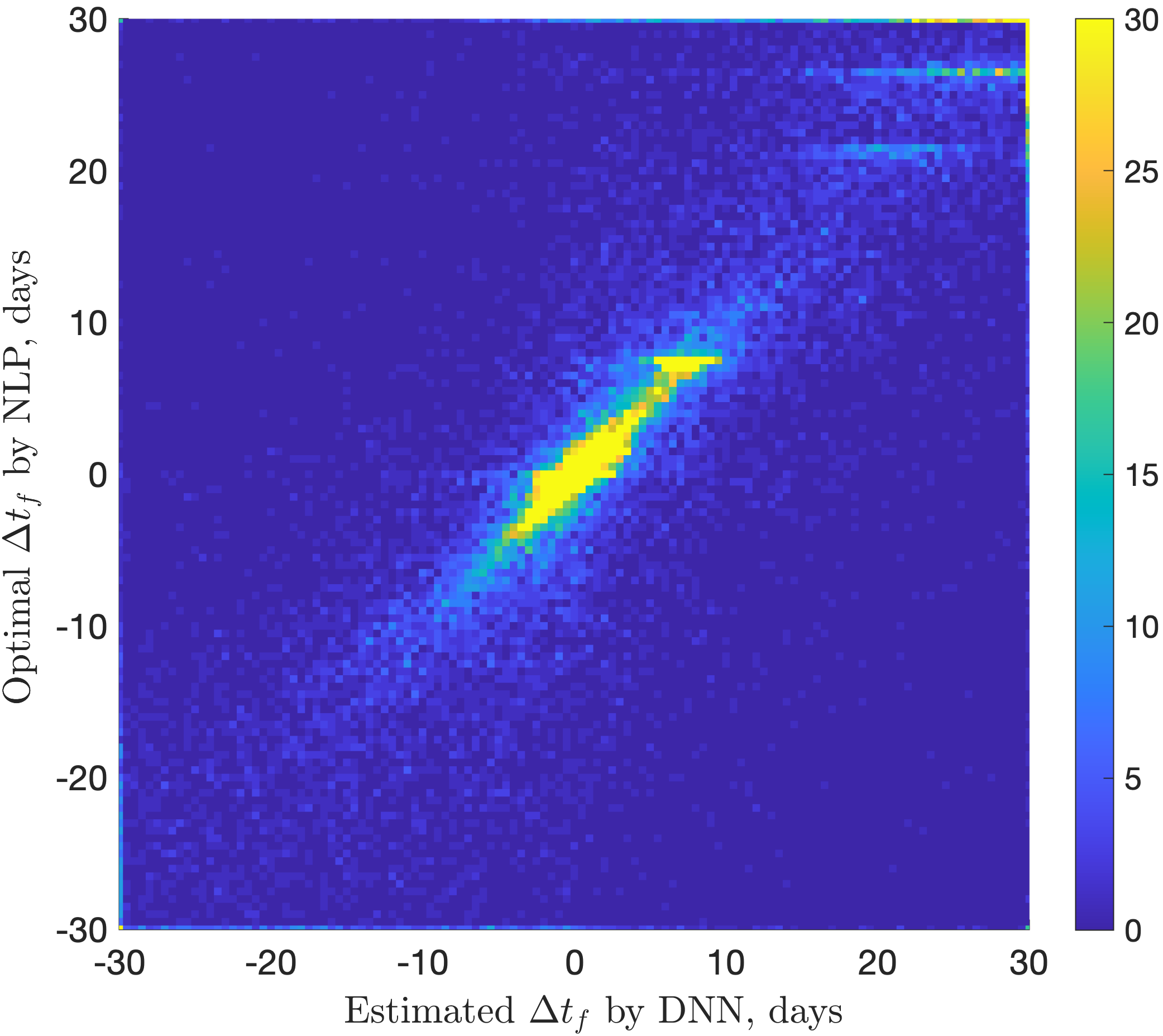}}%
\subfigure[Estimated $\Delta v_{\infty, f}$ vs true $\Delta v_{\infty, f}$]{%
    \includegraphics[clip, width=0.45\columnwidth]{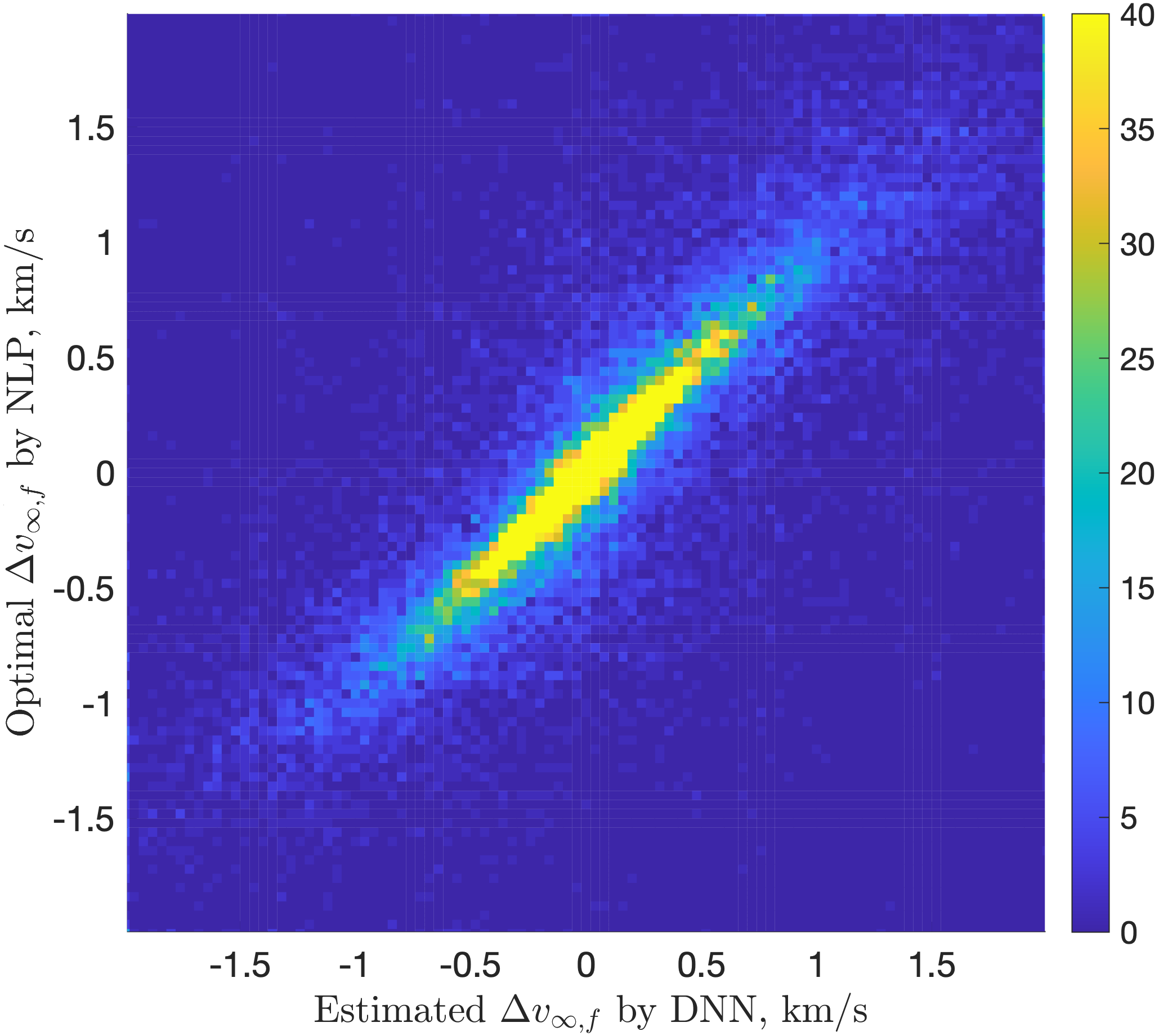}}%
\caption{\majorblue{Density heatmaps} of the estimated outputs by DNNs vs the true outputs (Color map: counts of the solution).}
\label{f:dtof_dvinf_compare}
\end{figure}

Figures \ref{f:dv_compare} and \ref{f:dtof_dvinf_compare} plot \majorblue{density heatmaps} that illustrate the correlation between the estimated values and the true values. These \majorblue{density heatmaps} are evaluated using 25,461 test trajectories \majorblue{(without pseudo-asteroids)} that are independent of the DNNs training and validation process. Figure \ref{f:dv_compare} compares the $\Delta V$ prediction performance of the DNN surrogate with Lambert's solution, typically used to estimate the optimal $\Delta V$. The estimated $\Delta V$ by DNN and the true $\Delta V$ have a strong 1-to-1 correlation, whereas the Lambert's $\Delta V$ and the true $\Delta V$ have a weak correlation. The $\Delta V$ error magnitude of the DNN surrogates is \majorblue{about 0.1km/s or less}, which is significantly improved from the typical Lambert's solution, as shown in Fig.\ref{f:dv_compare} (b). Figure \ref{f:dtof_dvinf_compare} compares the DNN prediction performances of other outputs $\Delta t_f$ and $\Delta v_{\infty,f}$. We also observe strong 1-to-1 correlations for both predictions.

%
%
\subsection{Tree Search}

Integrating the Earth-asteroid-Earth block with the DNN surrogate model, we search for good asteroid flyby sequences via beam search. The parameters to be searched in the beam search are the type of free-return trajectories and the orbital elements of the asteroids. The ephemerides of the target asteroids are acquired via the JPL small-body database search engine\footnote{\url{https://ssd.jpl.nasa.gov/sbdb_query.cgi} (Accessed on July 22, 2021.)} under the condition of the perihelion radius $q\leq1.4$ au, the aphelion radius $Q\geq0.8$ au, and the OCC (Orbit Condition Code)$\leq6$. As of July 22, 2021, the number of the target asteroids is 15,340. 

Each parent has about 300,000 children in the tree search, and about 10,000 of them remain after the initial screening. The DNN surrogate evaluates the cost of 10,000 Earth-asteroid-Earth trajectories in 10 seconds, whereas the trajectory optimization takes about 1140 seconds for the same computation. Suppose that the beam width is 100 and the maximum depth is 5, the total computation time of our algorithm is about 10 hours, including initial screening and all pre/post-processing of database, whereas the trajectory optimization (without the DNN surrogate) is expected to take about 7 days for the same computation.

%
%
\subsection{Patched Trajectory Optimization}

\majorred{Using the flyby sequence provided by the proposed method, we solve a multiple gravity-assist trajectory optimization problem that patches the whole trajectory. We patch each Earth-asteroid-Earth trajectory with an Earth gravity assist, modeled by a zero-radius sphere-of-influence patched-conics approach (e.g., \cite{Vasile2006, Ozaki2022a}). The gravity assist maneuver is modeled as an instantaneous velocity change with constraints on the hyperbolic excess velocity, deflection angle, and epoch of the Earth flyby between phases.} We use the results of the DNN surrogate and the screening algorithms as the initial guess trajectories. This patched trajectory optimization was performed via a direct multiple shooting method that minimizes the total $\Delta V$ magnitude. The initial \majorblue{hyperbolic excess velocity} and the Earth departure time are bounded with the $\pm$0.2 km/s and $\pm$7 days tolerance.



Table \ref{tab:flyby_cycler_examples} shows some of the \majorblue{example} patched asteroid flyby cycler trajectories. We compare the total $\Delta V$s between the DNN-based method and NLP-based patched trajectory optimization. The total $\Delta V$ of NLP differs from the predicted $\Delta V$ of DNNs because of the following factors. \majorred{1) Prediction errors of DNNs result in errors in $\Delta V$, Earth gravity assist epoch, and hyperbolic excess velocity, as shown in Fig.\ref{f:dv_compare} (a), Fig.\ref{f:dtof_dvinf_compare} (a), and Fig.\ref{f:dtof_dvinf_compare} (b), respectively. Because the errors on the Earth gravity assist epoch and hyperbolic excess velocity change the boundary condition of the Earth-asteroid-Earth block, these errors are accumulated in the total $\Delta V$ for the patched trajectory optimization. For reference, as illustrated in Fig.\ref{f:dv_compare} (a), the $\Delta V$ error of the single-stage DNN surrogates ($=$Optimal $\Delta V$ by NLP $-$ Estimated $\Delta V$ by DNN) is about 0.1km/s or less in most cases. The large error seen in ID1-0704 in Table \ref{tab:flyby_cycler_examples} is likely because of this accumulation of errors in the Earth gravity assist epoch and hyperbolic excess velocity.} 2) The patched trajectory optimization reduces the total $\Delta V$ consumption \majorred{because the patched trajectory optimization solves the whole trajectory whereas the initial guess trajectory provided by DNNs is a sequence of optimal solutions to the subproblem in each Earth-asteroid-Earth block.} 


\begin{table*}[htb]
    \begin{center}
        \caption{List of \majorblue{example} patched asteroid flyby cycler trajectories}
        \label{tab:flyby_cycler_examples}
        \scalebox{0.9} {
        \begin{tabular}{cccccccc}
            \hline
            \multirow{2}{*}{ID} & \multicolumn{2}{c}{Total $\Delta V$, km/s} & \multicolumn{2}{c}{Total TOF, year} & \multicolumn{2}{c}{Free-return info} & Target body info \\
            & DNN & NLP & DNN & NLP & Leg & Type (m:n, $V_{\infty, \textrm{EGA0}}$ km/s) & Name (H, OCC, PHA, SMASSII spec.)\\\hline
            & & & & & 1 & full (2:2, 2.706) & 2000 MU1 (19.88, 0, Y, S)  \\
            & & & & & 2 & full (2:2, 2.706) & 2005 VQ96 (20.4, 0, Y, -)\\
            2-3094 & 0.1850 & 0.05101 & 9.488 & 9.507 & 3 & full (2:2, 2.708) & 2019 RC2 (22.5, 6, N, -)\\
            & & & & & 4 & full (2:2, 2.714) & 2016 GU (25.7, 6, N, -)\\
            & & & & & 5 & generic (1:1, 2.719) & 2000 AC6 (21.5, 0, Y, Q)\\\hline 
            & & & & & 1 & full (1:1, 2.764) & 2017 AE5 (22.3, 0, N, -)  \\
            & & & & & 2 & full (1:1, 2.764) & 2014 UU56 (28.6, 6, N, -)\\
            2-0866 & 0.09012 & 0.06208 & 5.993 & 5.996 & 3 & full (1:1, 2.788) & 2016 QE45 (21.78, 0, Y, -)\\
            & & & & & 4 & full (2:2, 2.751) & 2013 CY (28.3, 5, N, -)\\
            & & & & & 5 & full (1:1, 2.751) & Vishnu (18.32, 0, Y, O)\\\hline 
            & & & & & 1 & full (1:1, 2.567) & 2017 YV8 (27.3, 5, N, -)  \\
            & & & & & 2 & full (2:2, 2.567) & 2021 BA (26.01, 6, N, -)\\
            2-8373 & 0.09739 & 0.06446 & 9.488 & 9.479 & 3 & full (2:2, 2.543) & 1989 UQ (19.5, 0, Y, B)\\
            & & & & & 4 & full (2:2, 2.543) & 2017 UX5 (19.8, 2, Y, -)\\
            & & & & & 5 & generic (2:2, 2.545) & 1988 XB (17.96, 0, Y, B)\\\hline 
            & & & & & 1 & full (2:2, 2.535) & 2017 UX5 (19.8, 2, Y, -)  \\
            & & & & & 2 & full (2:2, 2.536) & 1999 MN (21.15, 0, Y, -)\\
            2-4497 & 0.1996 & 0.1139 & 9.014 & 9.010 & 3 & full (2:2, 2.550) & 2017 UX5 (19.8, 2, Y, -)\\
            & & & & & 4 & full (2:2, 2.550) & 2021 FC (28.25, 4, N, -)\\
            & & & & & 5 & full (1:1, 2.442) & 2013 WT67 (17.98, 0, Y, U)\\\hline 
            & & & & & 1 & half (1.5:1.5, 2.691) & 1998 XX2 (19.9, 0, Y, -)  \\
            & & & & & 2 & full (1:1, 2.689) & 2003 LN6 (24.6, 2, N, -)\\
            1-1798 & 0.2101 & 0.1242 & 5.956 & 5.944 & 3 & full (1:1, 2.689) & 2016 JJ17 (22.9, 3, N, -)\\
            & & & & & 4 & full (1:1, 2.689) & 2005 QP11 (26.4, 4, N, -)\\
            & & & & & 5 & generic (1:1, 2.717) & 2000 WO107 (19.28, 0, Y, X)\\\hline 
            & & & & & 1 & full (1:1, 2.785) & 2017 AE5 (22.3, 0, N, -)  \\
            & & & & & 2 & full (1:1, 2.785) & 2014 UU56 (28.6, 6, N, -)\\
            1-0160 & 0.1682 & 0.1655 & 5.994 & 5.997 & 3 & half (1.5:1.5, 2.745) & 2000 WF3 (23.4, 4, N, -)\\
            & & & & & 4 & full (1:1, 2.734) & 2020 OB6 (24.9, 5, N, -)\\
            & & & & & 5 & half (1.5:1.5, 2.734) & 1996 EN (16.39, 0, Y, U)\\\hline  
            & & & & & 1 & full (2:2, 2.551) & 2021 GB8 (27.36, 6, N, -)  \\
            & & & & & 2 & full (2:2, 2.551) & 2014 KG39 (25, 3, N, -)\\
            2-0587 & 0.1611 & 0.2589 & 9.477 & 9.511 & 3 & full (2:2, 2.551) & 2019 TE3 (25.5, 6, N, -)\\
            & & & & & 4 & full (2:2, 2.581) & 2017 KB3 (25.1, 6, N, -)\\
            & & & & & 5 & generic (1:1, 2.581) & P/2016 BA14 PANSTARRS (-, 1, -, -)\\\hline 
            & & & & & 1 & full (1:1, 2.798) & 2017 AE5 (22.3, 0, N, -)  \\
            & & & & & 2 & full (1:1, 2.798) & 2014 UU56 (28.6, 6, N, -)\\
            1-0704 & 0.08417 & 0.3204 & 4.992 & 4.998 & 3 & full (1:1, 2.755) & 2017 WX13 (24.4, 6, N, -)\\
            & & & & & 4 & full (1:1, 2.755) & 2006 CJ (20.2, 0, Y, -)\\
            & & & & & 5 & full (1:1, 2.733) & Midas (15.22, 0, Y, V)\\\hline 
        \end{tabular}
        }
    \end{center}
\end{table*}

%
%
\subsection{\majorblue{Example} Solutions}

This subsection shows the details of two \majorblue{example} solutions. Two cases, ID 2-8373 and ID 1-1798, are presented as follows.

\begin{figure}
\subfigure[Sun-centered, ECLIPJ2000 inertial frame]{%
    \includegraphics[clip, width=0.5\columnwidth]{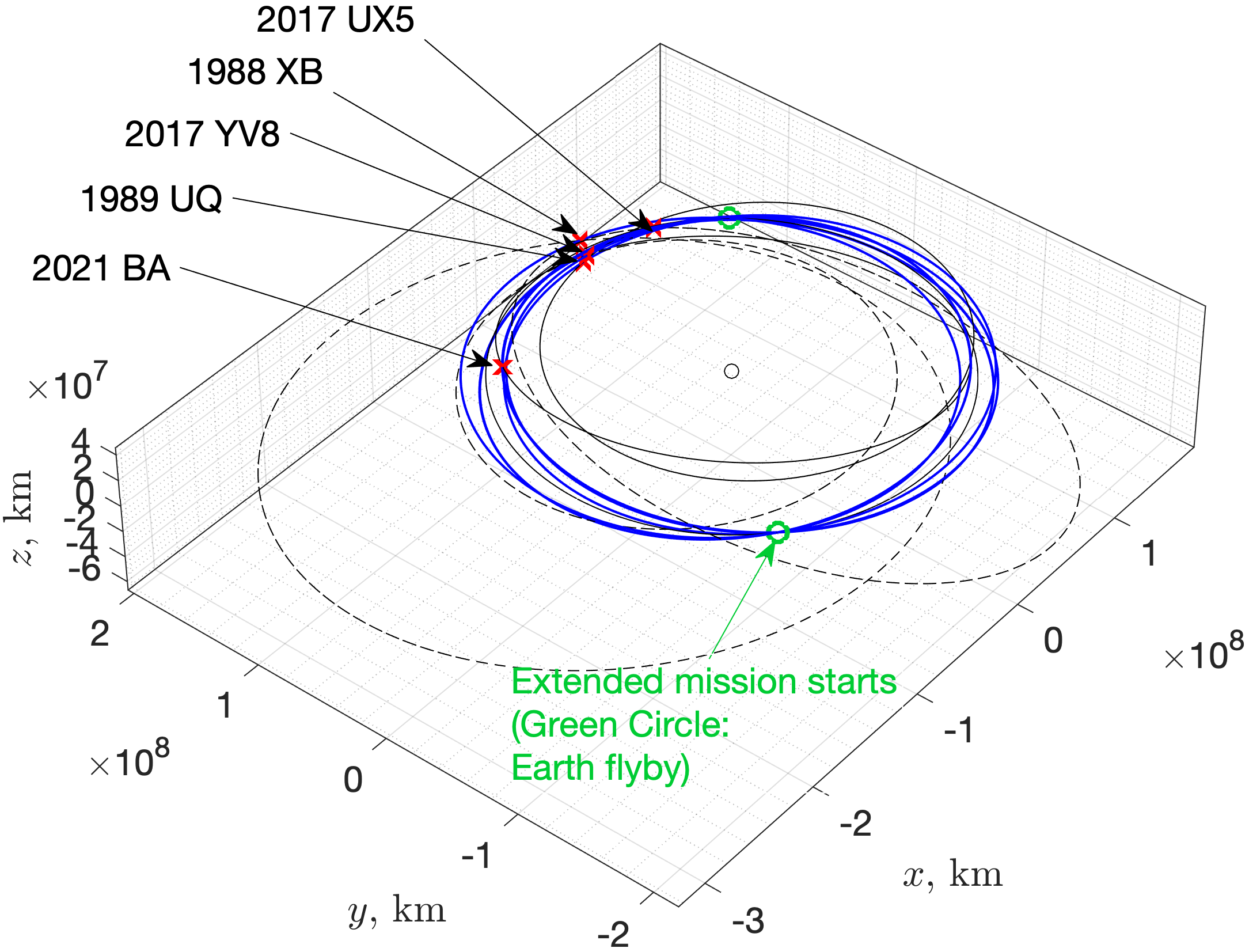}}%
\subfigure[Earth-centered, Sun-Earth line fixed rotational frame]{%
    \includegraphics[clip, width=0.5\columnwidth]{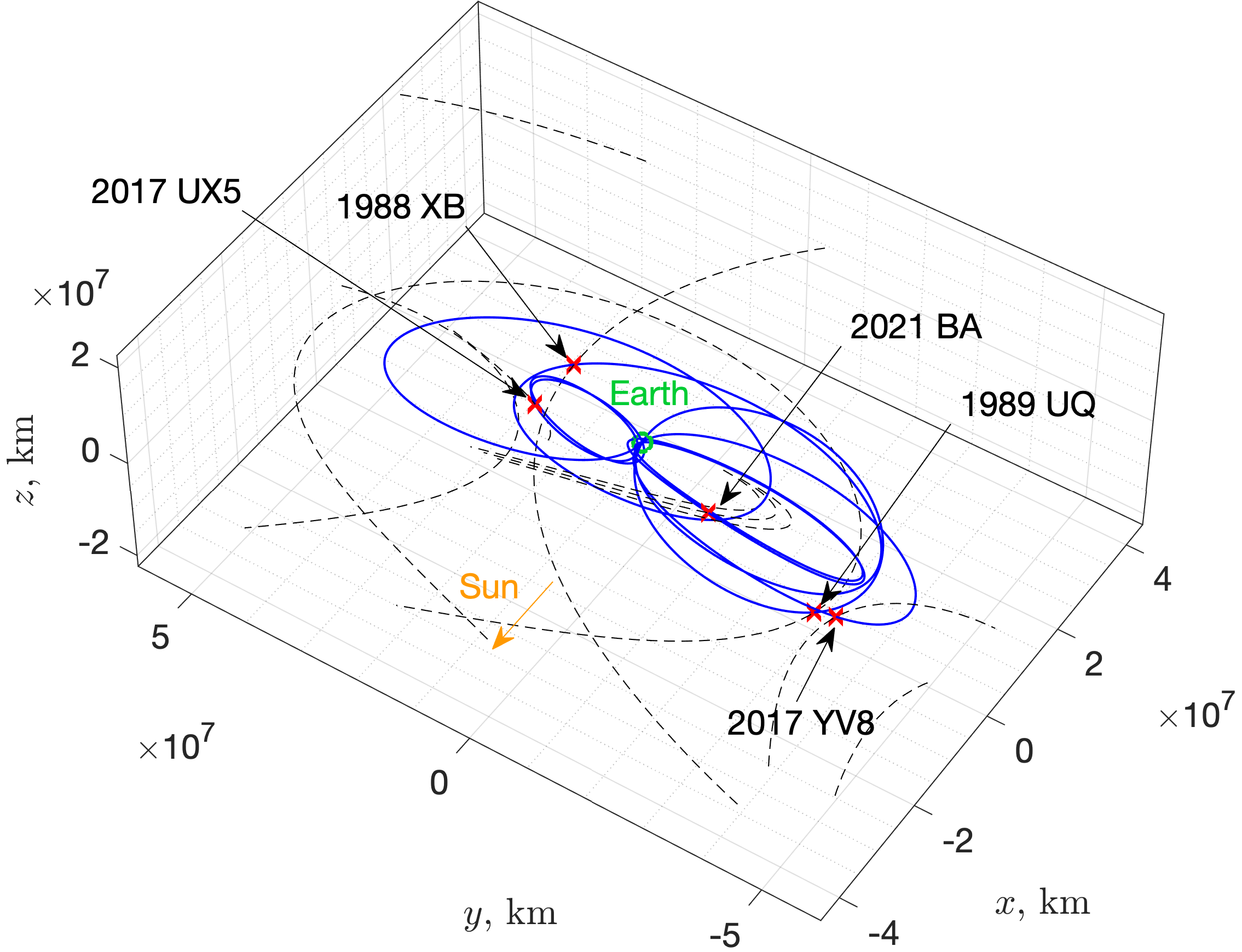}}%
\caption{ID 2-8373 asteroid flyby cycler trajectory (nearly ballistic, long transfer time)}
\label{f:id2_8373_trajectories}
\end{figure}

ID 2-8373 is a nearly ballistic solution that includes an exploration of two scientifically interesting B-type asteroids, which are thought to contain primitive and volatile-rich materials. Figure \ref{f:id2_8373_trajectories} shows the trajectories, and Table \ref{tab:flyby_table_2_8373} summarizes the sequence of events. The spacecraft can fly by an asteroid every few years. The total $\Delta V$ of 65 m/s is affordable for micro spacecraft and CubeSats, such as PROCYON\cite{Campagnola2015, Funase2015} and EQUULEUS\cite{Campagnola2019eq}.


\begin{table}[H]
\caption{\label{tab:flyby_table_2_8373} Sequence of events for ID 2-8373} 
\centering
\begin{tabular}{lllll}
\hline
Date time, TDB & Event & $v_{\infty}$ (or $v_{\textrm{rel}}$), km/s & Perigee altitude, km & $\Delta V$, km/s\\\hline
2028 MAY 12 00:52:51 & Earth flyby & 2.567 & 232385 & \\
2028 DEC 27 10:50:45 & 2017 YV8 flyby & 10.063 &  &  \\
2029 MAY 12 07:05:22 & Earth flyby & 2.567 & 232386 & \\
2030 FEB 10 00:45:50 & 2021 BA flyby & 9.453 &  &  \\
2030 JUN 13 16:53:49 & Deep space maneuver \#1 & & & 0.0285 \\
2031 MAY 12 14:49:00 & Earth flyby & 2.543 & 40380 & \\
2031 DEC 26 19:28:45 & 1989 UQ flyby & 6.501 &  &  \\
2033 MAY 12 03:13:28 & Earth flyby & 2.543 & 29185 & \\
2033 NOV 12 10:58:02 & 2017 UX5 flyby & 13.859 &  &  \\
2034 MAY 17 14:53:20 & Deep space maneuver \#2 & & & 0.0219 \\
2035 MAY 12 15:40:17 & Earth flyby & 2.545 & 37295 & \\
2036 JAN 30 23:48:10 & Deep space maneuver \#3 & & & 0.0141\\
2036 DEC 01 07:20:19 & 1988 XB flyby & 11.423 &  &  \\
2037 OCT 31 20:46:55 & Earth flyby & 2.545 & n/a & \\
\hline
\end{tabular}
\end{table}

\begin{figure}[H]
\subfigure[Sun-centered, ECLIPJ2000 inertial frame]{%
    \includegraphics[clip, width=0.5\columnwidth]{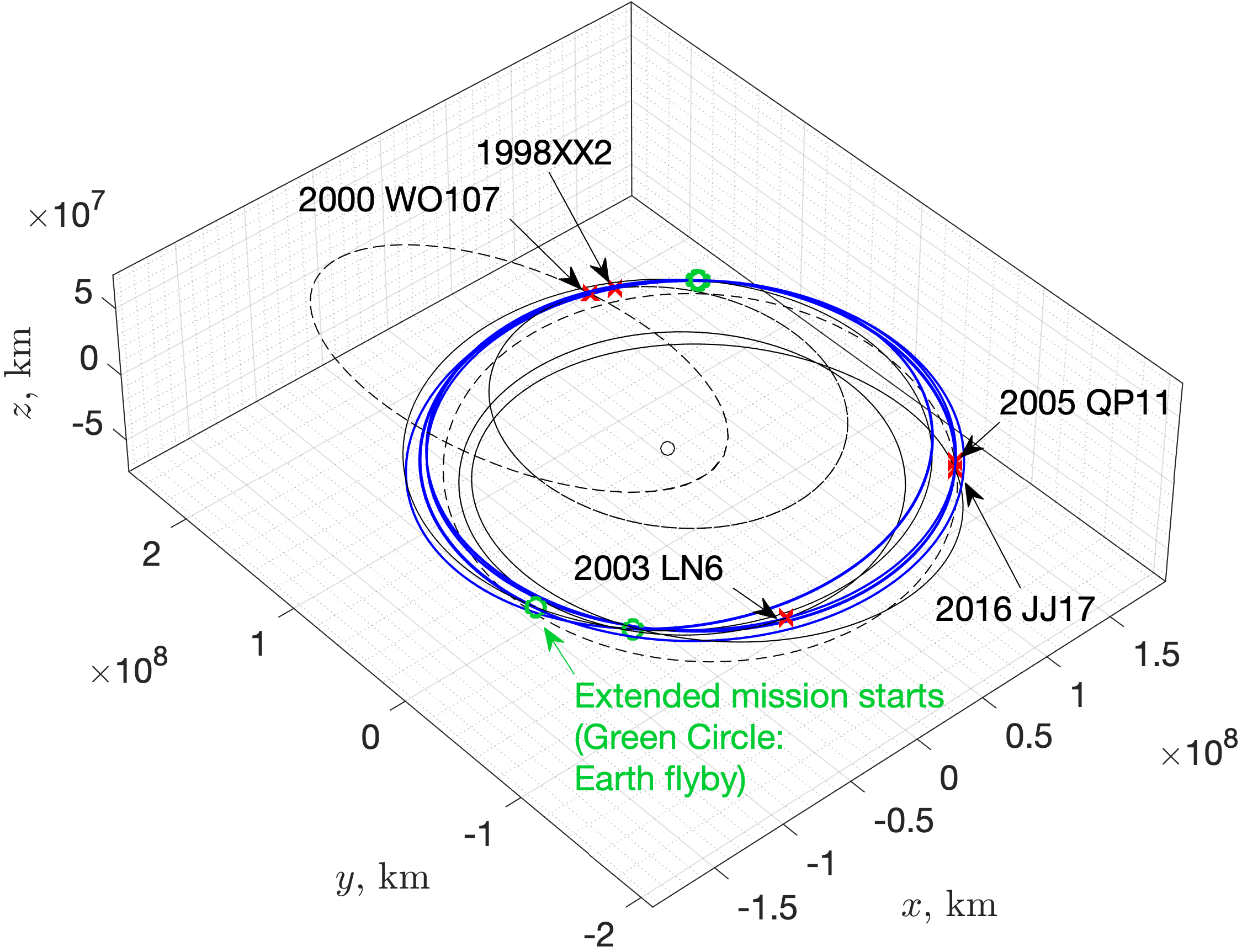}}%
\subfigure[Earth-centered, Sun-Earth line fixed rotational frame]{%
    \includegraphics[clip, width=0.5\columnwidth]{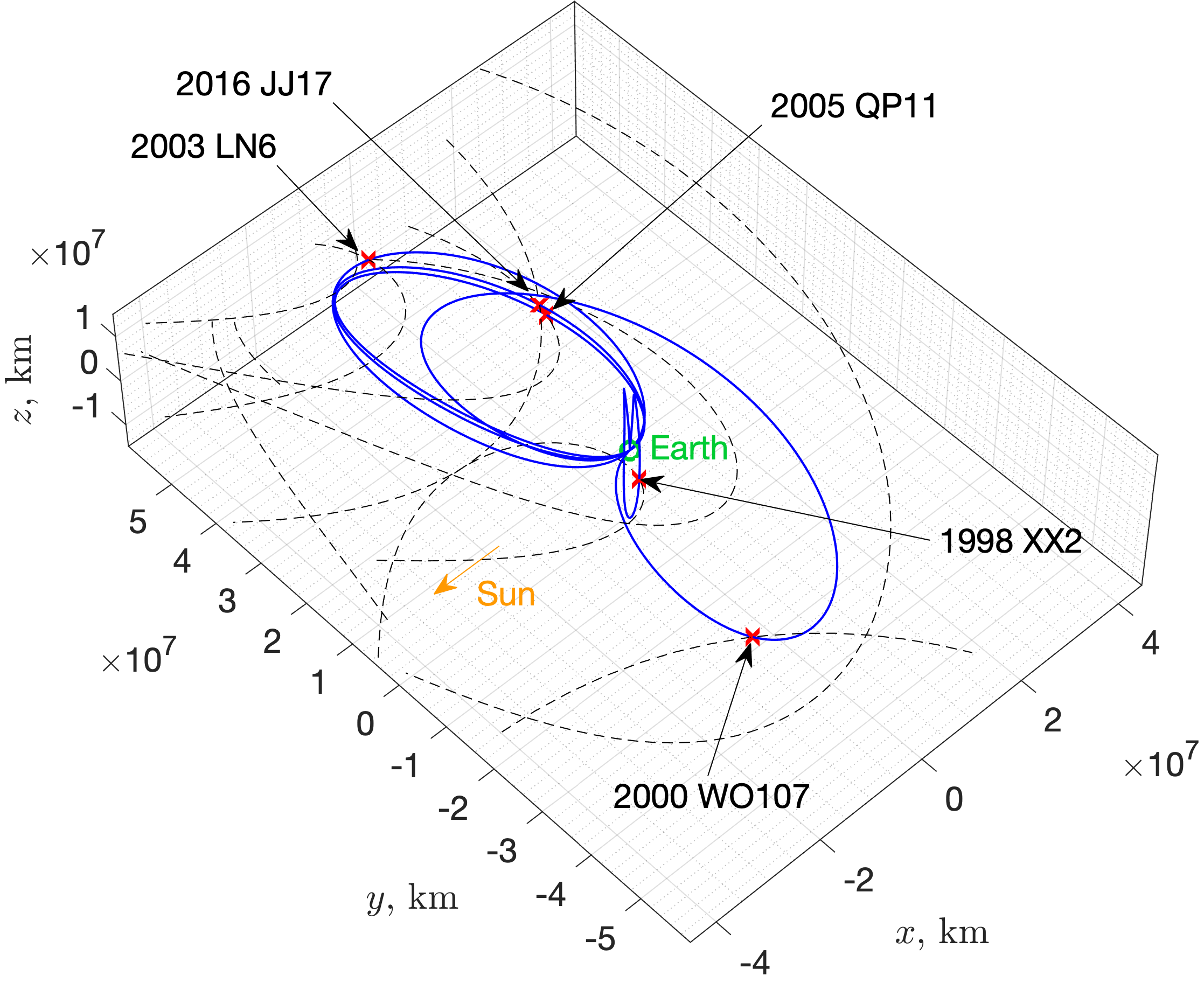}}%
\caption{ID 1-1798 asteroid flyby cycler trajectory (short transfer time)}
\label{f:id1_1798_trajectories}
\end{figure}

ID 1-1798 is a short time transfer solution that includes an exploration of 2000 WO107, which is a contact binary asteroid. Such an irregular-shaped asteroid is important to understand the collisional environment in the early solar system. Figure \ref{f:id1_1798_trajectories} shows the trajectories, and Table \ref{tab:flyby_table_1_1798} summarizes the sequence of events. The spacecraft can fly by an asteroid almost every year. The total $\Delta V$ of 124 m/s is still reasonable for many missions.

\begin{table}[H]
\caption{\label{tab:flyby_table_1_1798} Sequence of events for ID 1-1798} 
\centering
\begin{tabular}{lllll}
\hline
Date time, TDB & Event & $v_{\infty}$ (or $v_{\textrm{rel}}$), km/s & Perigee altitude, km & $\Delta V$, km/s\\\hline
2028 MAY 04 12:08:25 & Earth flyby & 2.691 & 11984 & \\
2028 NOV 19 01:59:41 & Deep space maneuver \#1 & & & 0.0228 \\
2028 NOV 23 07:49:22 & 1998 XX2 flyby & 8.545 &  &  \\
2029 MAY 13 22:38:40 & Deep space maneuver \#2 & & & 0.0010 \\
2029 NOV 03 09:26:10 & Earth flyby & 2.689 & 24795 & \\
2030 MAY 17 15:23:49 & 2003 LN6 flyby & 4.180 &  &  \\
2030 NOV 03 15:38:28 & Earth flyby & 2.689 & 297689 & \\
2031 JUL 26 22:13:06 & 2016 JJ17 flyby & 8.369 &  &  \\
2031 NOV 03 21:50:36 & Earth flyby & 2.689 & 1180355 & \\
2032 FEB 05 07:41:40 & Deep space maneuver \#3 & & & 0.0169\\
2032 JUL 29 09:25:09 & 2005 QP11 flyby & 4.054 &  &  \\
2032 NOV 03 21:37:26 & Earth flyby & 2.717 & 245400 & \\
2033 AUG 11 07:23:50 & Deep space maneuver \#4 & & & 0.0835\\
2033 DEC 13 18:52:26 & 2000 WO107 flyby & 28.477 &  &  \\
2034 APR 13 04:44:34 & Earth flyby & 2.779 & n/a & \\
\hline
\end{tabular}
\end{table}

%
%
\section{Conclusions}

Asteroid flyby cycler orbits allow for multiple asteroid flybys with little $\Delta V$ consumption. This paper proposes a novel trajectory design approach of asteroid flyby cyclers utilizing the surrogate model via deep neural networks. \majorred{The proposed architecture creates the Earth-asteroid-Earth blocks integrating the surrogate model with astrodynamics knowledge, such as free-return trajectories and Lambert's problem, to improve the prediction accuracy. The resulting surrogate models are reusable for different mission scenarios (different Earth departure epoch, hyperbolic excess velocity, and target asteroids) without re-training deep neural networks. Because machine learning-based trajectory design requires a computationally expensive gigantic database, we propose an efficient database generation strategy that can amplify the size of the optimal trajectory database by one order of magnitude by introducing pseudo-asteroids satisfying the Karush-Kuhn-Tucker conditions. These surrogate-based Earth-asteroid-Earth blocks allow us to search for good asteroid flyby sequences via beam search efficiently.} The numerical application to JAXA's \destiny mission, an upcoming asteroid flyby mission, shows that the proposed method \majorblue{is practically applicable to space mission design and} efficiently finds the asteroid flyby cycler trajectories.


%
\section*{Acknowledgments}\label{Acknowledgments}

This research is supported by the Adaptable and Seamless Technology Transfer Program (A-STEP) through Target-driven R\&D, Grant Number JPMJTM20D9, from Japan Science and Technology Agency (JST). The first author would like to thank JAXA's \destiny project team for their valuable comments.

%
%
%
%
\section*{Appendix A. Detail of Inputs of Deep Neural Network}
\renewcommand{\theequation}{A.\arabic{equation} }
\setcounter{equation}{0}

This section explains the details of the input information of DNNs shown in Tables \ref{tab:lambert_dnn} and \ref{tab:distance_dnn}. Free-return info $(m, n, \textrm{type}, v_{\infty,0})$ and asteroid orbital elements $(a, e, i, \Delta \Omega_{t_0},\omega, M_{t_0})$ are user-defined parameters, which are parametrically searched via tree search methods.

\majorblue{$T$} is the time of flight of the free-return trajectory. $n_{\star}$ is the number of revolutions of the spacecraft's orbit from Earth departure to the asteroid encounter. $\eta_{t{\star}}$ is a parameter that determines the asteroid flyby epoch by the following equation.
\begin{equation}
    \eta_{t_{\star}} = \frac{t_1-t_0}{\majorblue{T}}.
\end{equation}
$\bm{v}_{\infty,0}$, $\bm{v}_{\infty,f}$ and $\bm{v}_{\textrm{rel},1}$ are determined in the RSW frame with respect to the flyby body; the R axis is parallel to the position vector of the flyby body; the W axis is normal to the orbital plane of the flyby body; the S axis completes the right-handed system; $\textbf{\oe}_{\textrm{FR}}$ is the orbital element of the free-return trajectory calculated from $(m, n, \textrm{type}, v_{\infty,0})$.

We calculate $\Delta V$s of Lambert's method as
\begin{align}
    \Delta v_{0} &= \|\bm{v}_{\infty\textrm{out},0}\| - v_{\infty,0}\\
    \Delta v_{1} &= \|\bm{v}_{\textrm{rel,out},1}-\bm{v}_{\textrm{rel,in},1}\|\\
    \Delta v_{\textrm{total}} &= \Delta v_0 + \Delta v_1,
\end{align}

and the closest approach state differences as
\begin{align}
    \delta \bm{r}_{\textrm{CA}} &= R_z\left(-\lambda_{\oplus}(t_0)\right) \left( \bm{r}_{\textrm{sc},1} - \bm{r}_{\star} (t_1) \right)\\
    \delta \bm{v}_{\textrm{CA}} &= R_z\left(-\lambda_{\oplus}(t_0)\right) \left( \bm{v}_{\textrm{sc},1} - \bm{v}_{\star} (t_1) \right)\\
    R_z(\theta) &= \begin{bmatrix}
    \cos\theta & -\sin\theta & 0\\
    \sin\theta & \cos\theta & 0\\
    0 & 0 & 1
    \end{bmatrix}.
\end{align}

\section*{Appendix B. Learning Curves for All Cases}
\renewcommand{\theequation}{B.\arabic{equation} }
\setcounter{equation}{0}

\begin{figure}[H]
\subfigure[Total validation loss]{%
    \includegraphics[clip, width=0.5\columnwidth]{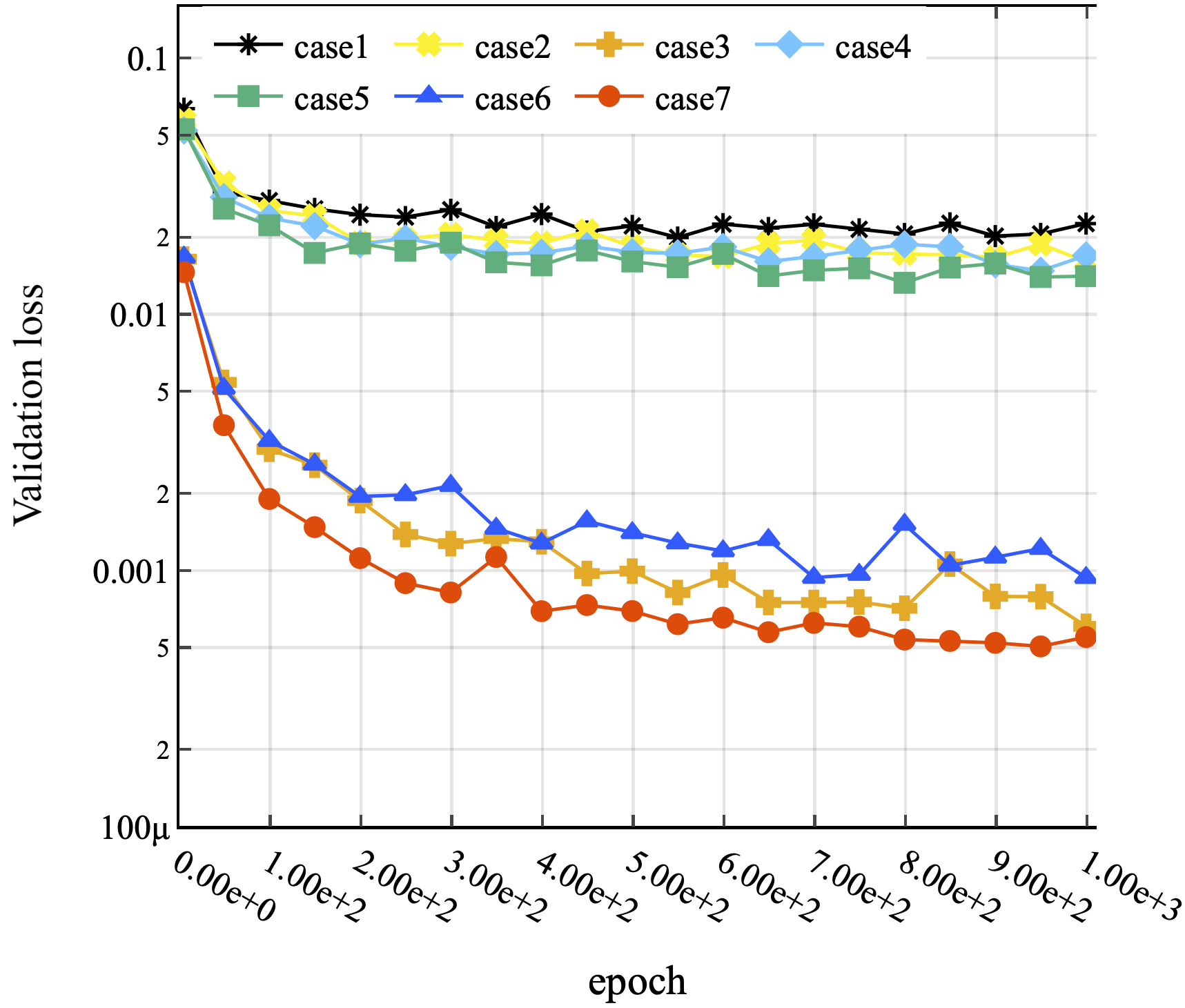}}%
\subfigure[Validation loss of $\Delta t_f$]{%
    \includegraphics[clip, width=0.5\columnwidth]{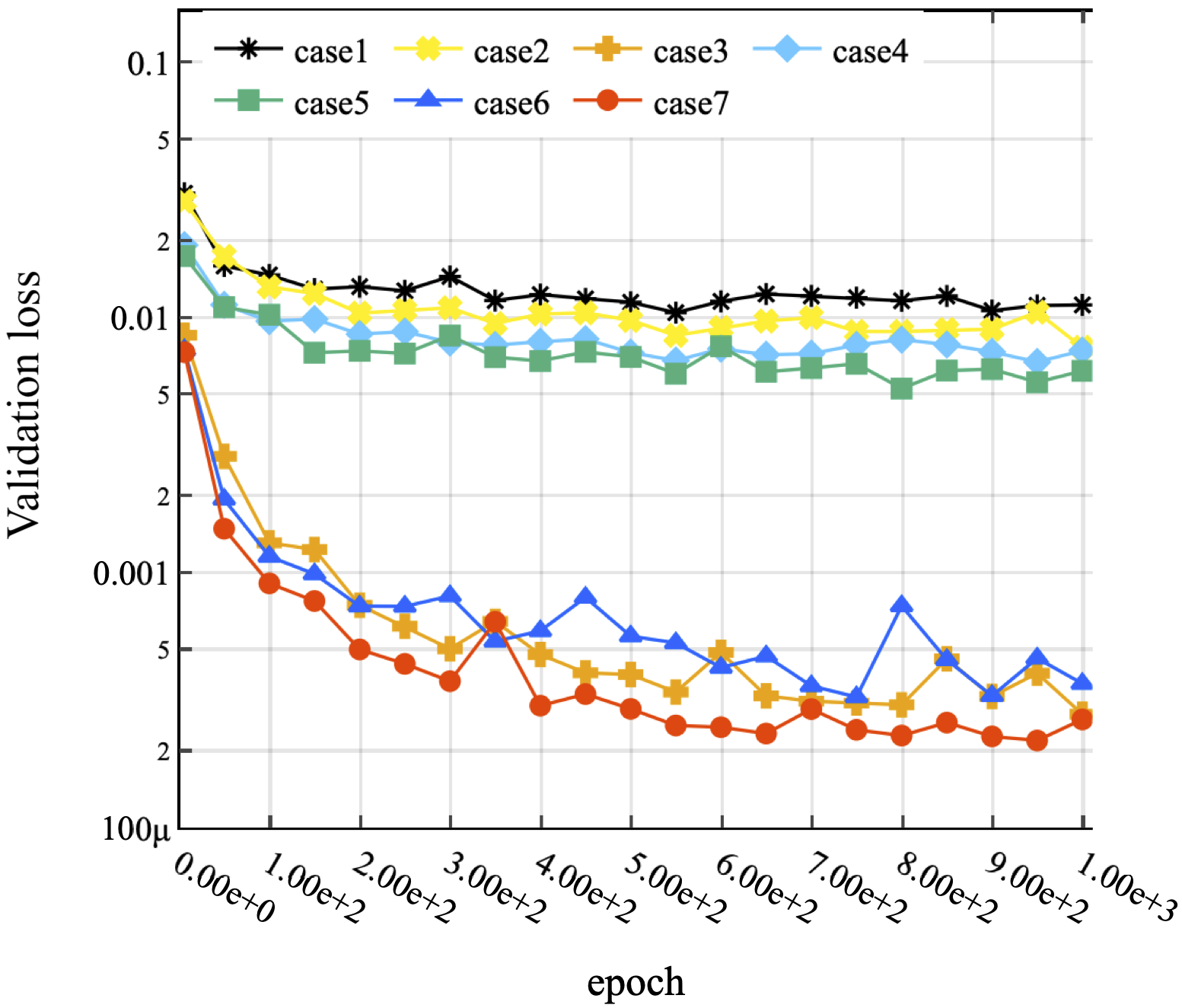}}%
    \\
\subfigure[Validation loss of $\Delta V$]{%
    \includegraphics[clip, width=0.5\columnwidth]{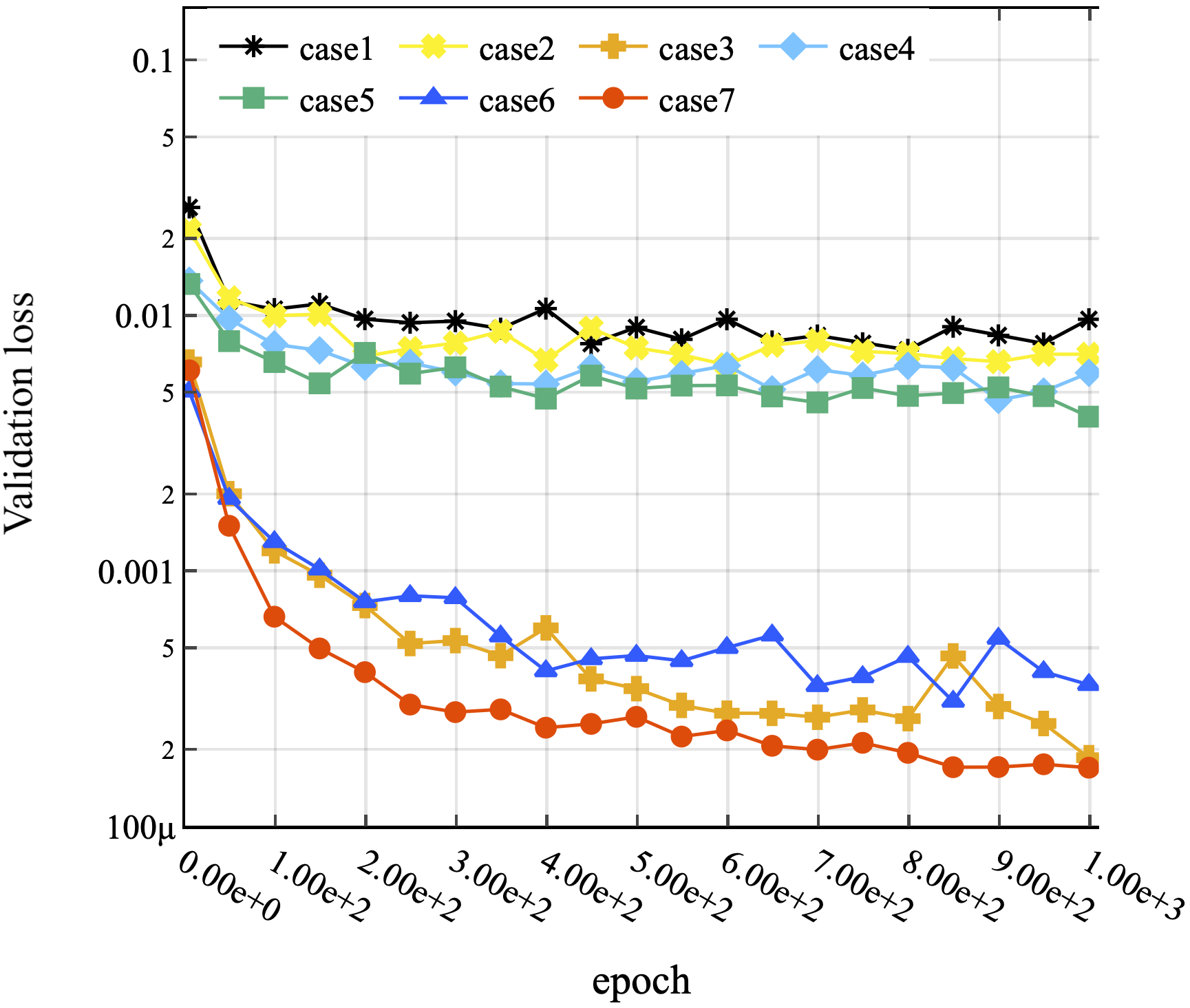}}%
\subfigure[Validation loss of $\Delta v_{\infty,f}$]{%
    \includegraphics[clip, width=0.5\columnwidth]{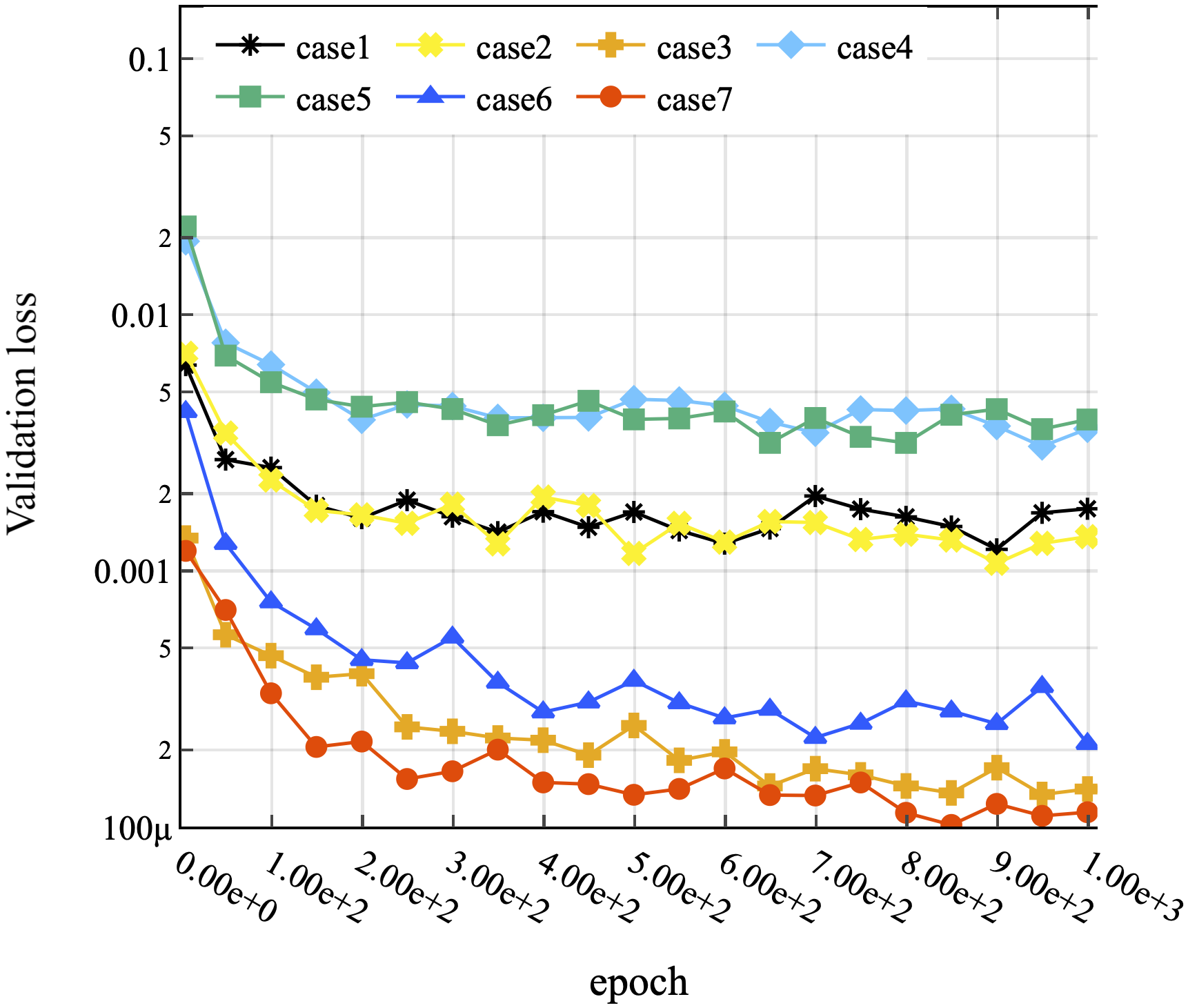}}%
\caption{Learning curve for all cases.}
\label{f:learning_curves}
\end{figure}

Figures \ref{f:learning_curves} illustrate the learning curves until 1k epoch for all cases. The definition of the corresponding database is written in Table \ref{tab:cases_database}. Case 7 records the best performance in all elements of the loss function.

\section*{\majorred{Appendix C. Sensitivity Analysis of Hyperparameters}}
\renewcommand{\theequation}{C.\arabic{equation} }
\setcounter{equation}{0}

\begin{figure}[H]
\subfigure[Validation loss for each \# of layers (\# of units is 1024 and learning rate is 1e-4)]{%
    \includegraphics[clip, width=0.5\columnwidth]{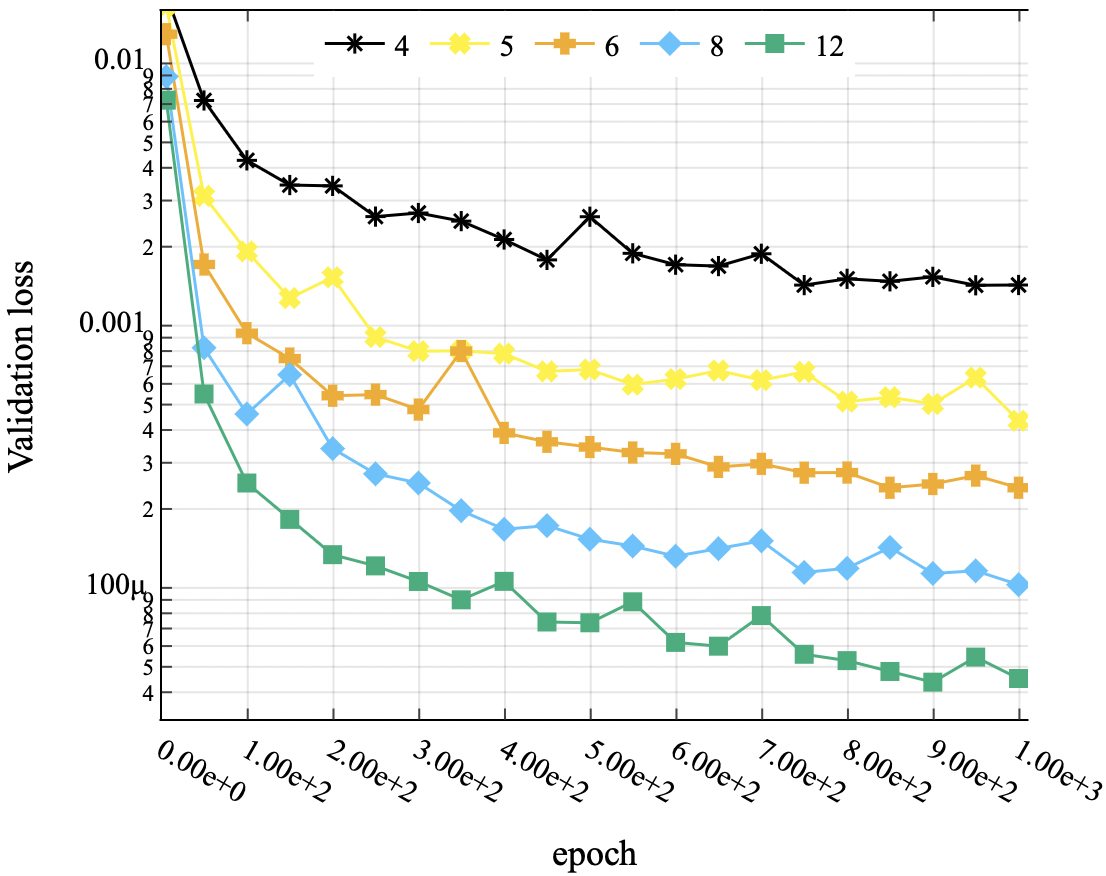}}%
\subfigure[Validation loss for each \# of units (\# of layers is 5 and learning rate is 1e-4)]{%
    \includegraphics[clip, width=0.5\columnwidth]{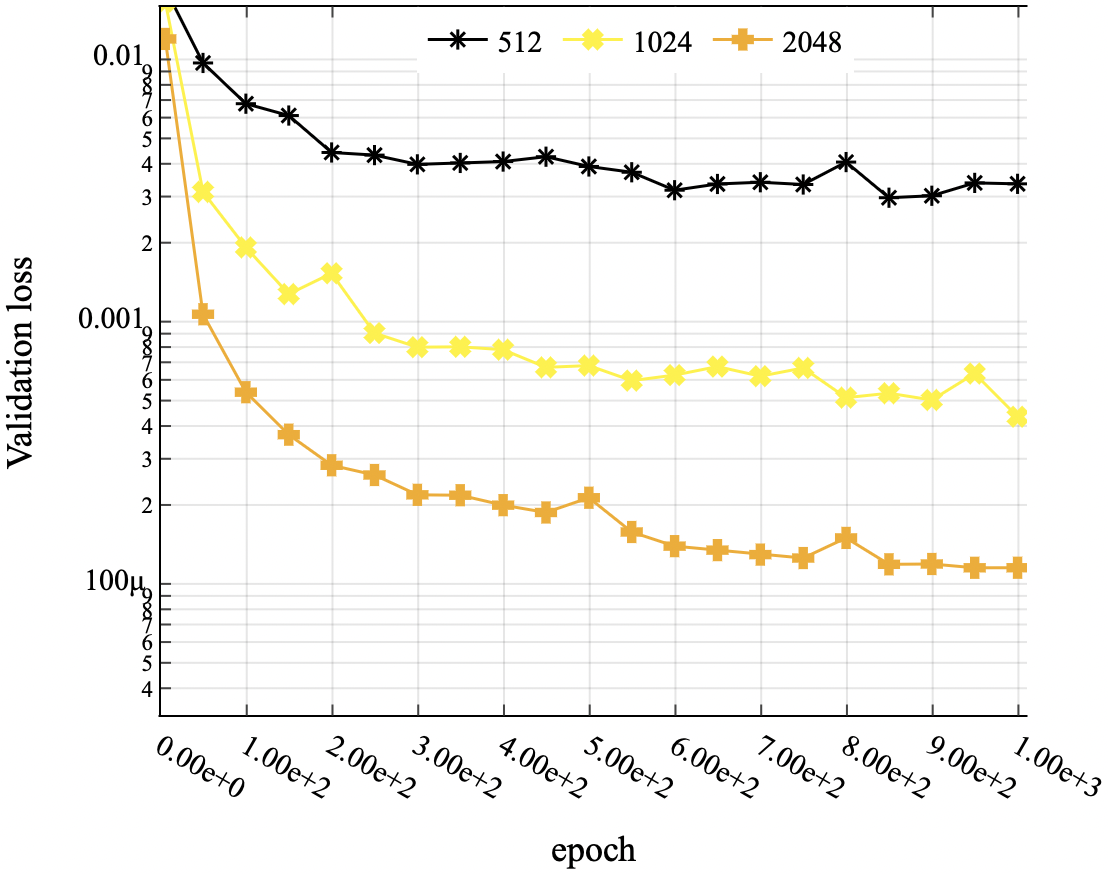}}%
    \\
\subfigure[Validation loss for each learning rate (\# of layers is 5 and \# of units is 1024)]{%
    \includegraphics[clip, width=0.5\columnwidth]{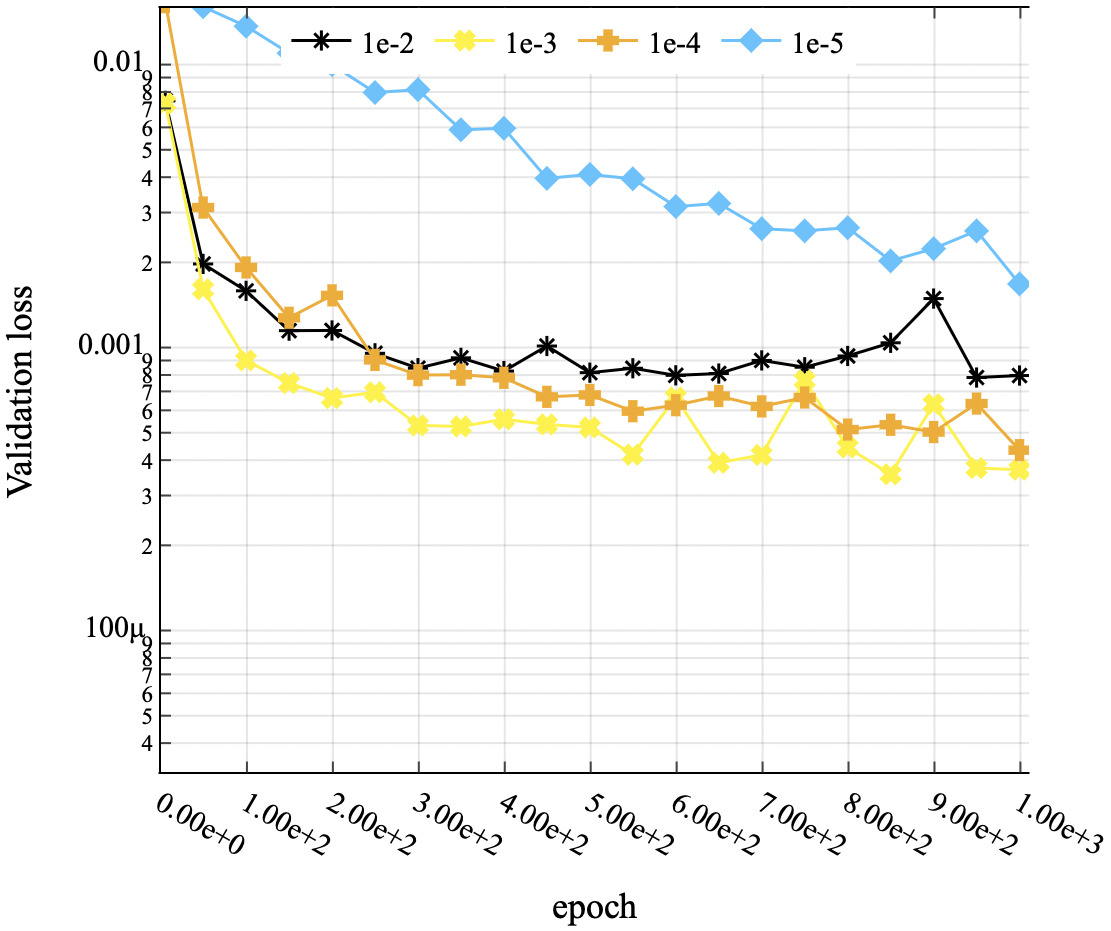}}%
\caption{Learning curve for each hyperparameters.}
\label{f:learning_curves_hyperparameters}
\end{figure}

\begin{figure}[H]
    \includegraphics[clip, width=0.45\columnwidth]{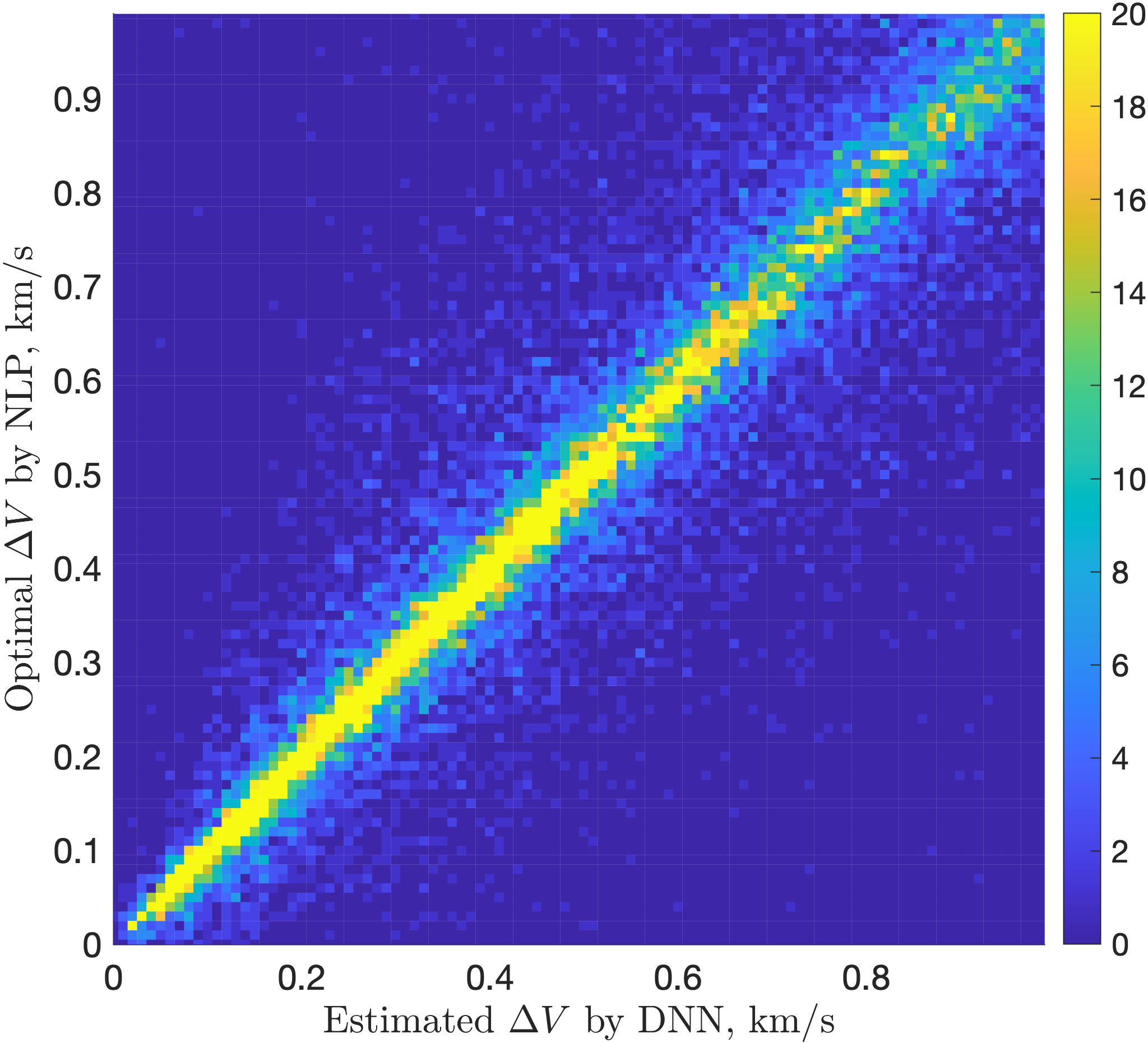}
    \caption{Density heatmaps of the estimated $\Delta V$s by DNNs vs the true optimal $\Delta V$s (Color map: counts of the solution).}
\label{f:dv_compare_new}
\end{figure}

\begin{figure}[H]
\subfigure[Estimated $\Delta t_f$ vs true $\Delta t_f$]{%
    \includegraphics[clip, width=0.45\columnwidth]{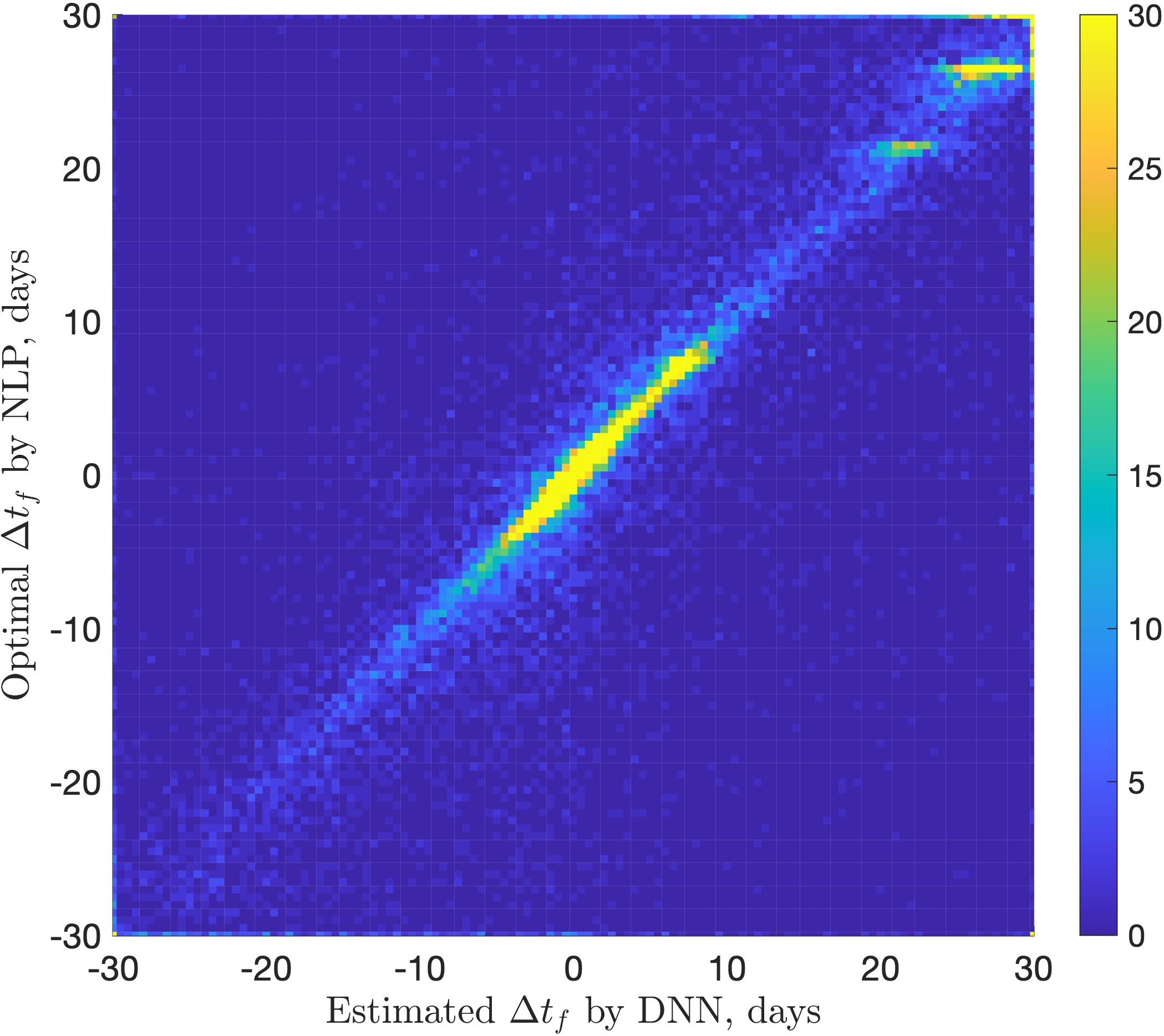}}%
\subfigure[Estimated $\Delta v_{\infty, f}$ vs true $\Delta v_{\infty, f}$]{%
    \includegraphics[clip, width=0.45\columnwidth]{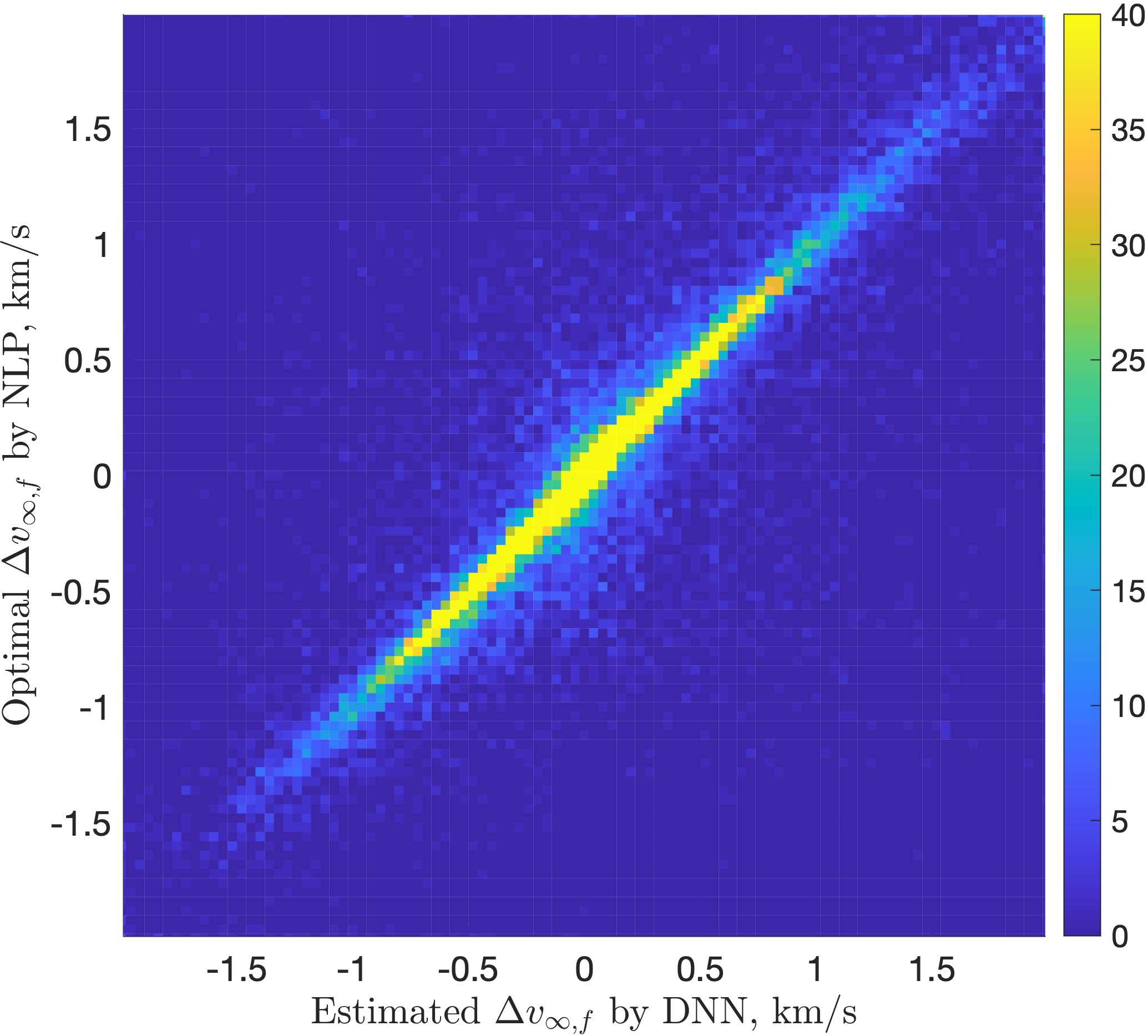}}%
\caption{Density heatmaps of the estimated outputs by DNNs vs the true outputs (Color map: counts of the solution).}
\label{f:dtof_dvinf_compare_new}
\end{figure}

\majorred{We perform sensitivity analysis of the hyperparameters in the database of Case 7. Figures \ref{f:learning_curves_hyperparameters} illustrate the learning curves until 1k epoch for each hyperparameters. The summary of the learning results is written in Table \ref{tab:hyperparameters_tuning}. The calculation results show that Case 7-j gives the optimal hyperparameters that minimize the validation loss. The density heatmaps evaluated using the test data are shown in Figs.\ref{f:dv_compare_new} and \ref{f:dtof_dvinf_compare_new}. Although we found better hyperparameters that decrease the validation loss, we selected Case 7-a, considering validation loss and calculation speed of training and predicting. As for the learning rate shown in Fig.\ref{f:learning_curves_hyperparameters} (c), we found that 1e-4 converges more stably than 1e-3, although the validation loss is slightly worse for 1e-4 than for 1e-3.}



\begin{table}[H]
\caption{\label{tab:hyperparameters_tuning} Performance for each hyperparameter}
\centering
\begin{tabular}{lccccc}
\hline
Case \# & \# of Layers & \# of Units & Learning rate & Validation loss (@epoch) & Speed (samples/s) \\\hline
7-a & 5 & 1024 & 1e-4 & 4.89e-4(@1k) & 195,000 \\ 
7-b & 5 & 1024 & 1e-3 & 3.69e-4(@1k) & 195,000 \\ 
7-c & 5 & 1024 & 1e-5 & 1.88e-3(@1k) & 195,000 \\ 
7-d & 6 & 1024 & 1e-4 & 2.53e-4(@1k) & 163,000 \\ 
7-e & 4 & 1024 & 1e-4 & 1.40e-3(@1k) & 261,000 \\ 
7-f & 5 & 2048 & 1e-4 & 1.24e-4(@1k) &  84,000 \\ 
7-g & 5 & 512 & 1e-4 & 3.36e-3(@1k) &  168,000 \\ 
7-h & 5 & 1024 & 1e-2 & 9.98e-4(@1k) & 195,000 \\ 
7-i & 8 & 1024 & 1e-4 & 1.10e-4(@1k) & 115,000 \\ 
7-j & 12 & 1024 & 1e-4 & 4.44e-5(@1k), 2.46e-5(@3.2k) & 74,000 \\ 
7-k & 6 & 2048 & 1e-4 & 5.76e-5(@1k) & 64,000 \\ 
\hline
\end{tabular}
\end{table}

%
%
\bibliography{references}   

\end{document}